\documentclass[twocolumn,english,prl,showpacs,preprintnumbers]{revtex4-1}
\usepackage[T1]{fontenc}
\usepackage[latin9]{inputenc}
\usepackage{textcomp}
\usepackage{amstext}
\usepackage{graphicx}
\usepackage{amssymb}

\makeatletter
\@ifundefined{textcolor}{}
{%
 \definecolor{BLACK}{gray}{0}
 \definecolor{WHITE}{gray}{1}
 \definecolor{RED}{rgb}{1,0,0}
 \definecolor{GREEN}{rgb}{0,1,0}
 \definecolor{BLUE}{rgb}{0,0,1}
 \definecolor{CYAN}{cmyk}{1,0,0,0}
 \definecolor{MAGENTA}{cmyk}{0,1,0,0}
 \definecolor{YELLOW}{cmyk}{0,0,1,0}
 }

\usepackage{dcolumn}\usepackage{bm}

\makeatother

\usepackage{babel}

\begin{document}

\title{Evidence for long-lived quasiparticles trapped in superconducting
point contacts}

\author{M.\ Zgirski}
\author{L.\ Bretheau}
\author{Q.\ Le Masne}
\author{H.\ Pothier}
\email[Corresponding author~: ]{hugues.pothier@cea.fr}
\author{D.\ Esteve}
\author{C.\ Urbina}

\affiliation{Quantronics Group, Service de Physique de l'État Condensé (CNRS,
URA\ 2464), IRAMIS, CEA-Saclay, 91191 Gif-sur-Yvette, France}

\date{\today}
\begin{abstract}
We have observed that the supercurrent across phase-biased, highly
transmitting atomic size contacts is strongly reduced within a broad
phase interval around $\pi$. We attribute this effect to quasiparticle
trapping in one of the discrete sub-gap Andreev bound states formed
at the contact. Trapping occurs essentially when the Andreev energy
is smaller than half the superconducting gap $\Delta,$ a situation
in which the lifetime of trapped quasiparticles is found to exceed
100~$\mu$s. The origin of this sharp energy threshold is presently
not understood.
\end{abstract}


\maketitle
Both theory and experiment indicate that the number of quasiparticles
in superconductors decreases exponentially as the temperature is lowered,
while their recombination time increases \cite{Kaplan,Wilson}. This
slow dynamics is an important ingredient in non-equilibrium superconductivity
and allows for the design of high-performance devices like single
photon detectors for astrophysical applications \cite{Day}. However,
recent developments on microwave resonators \cite{de Visser} and
Josephson qubits \cite{Lenander} show that at very low temperatures
residual non-equilibrium quasiparticles set a limit to the proper
functioning of these devices. More drastically, a single quasiparticle
can determine the response of single-Cooper pair devices \cite{Joyez}
containing small superconducting islands in which the parity of the
total number of electrons actually matters. The trapping of a single
quasiparticle in such a superconducting island has been dubbed {}``poisoning''
\cite{Aumentado}, as it inhibits the behavior expected in the ground
state of the system. Remarkably, it has been argued \cite{Chtchelkatchev}
that quasiparticle trapping could also occur in the discrete Andreev
bound states \cite{Furusaki} formed at sub-gap energies
in a constriction between two superconductors, a system containing
no island at all.%
\begin{figure}
\includegraphics[clip,width=1\columnwidth]{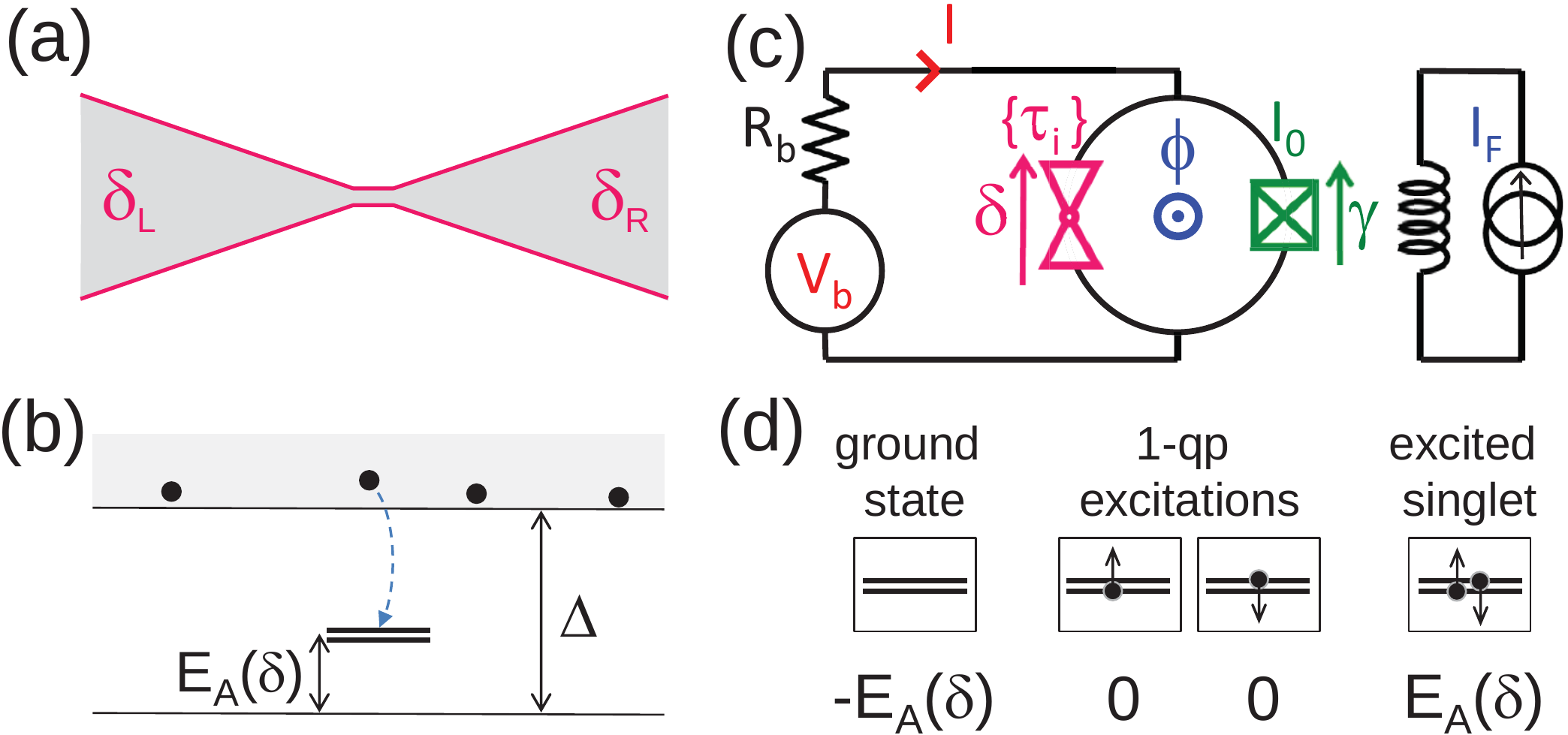}\caption{(Color online) \textit{(a)} Short one-channel constriction between
two superconducting electrodes (phase difference $\delta=\delta_{L}-\delta_{R}$).\emph{
}\textit{(b)} Excitation spectrum: besides the usual continuum of
states above the energy gap $\Delta$, that extends all across the
structure, there is at the constriction a discrete Andreev spin-degenerate
doublet with energy $E_{A}(\delta)$ above the ground state, where
quasiparticles can get trapped. \textit{(c)} Schematic setup: an atomic
point contact (red triangles) forms a SQUID with a Josephson junction
(green checked box). Phases $\delta$ and $\gamma$ across contact
and junction are linked by the flux $\phi$ threading the loop. The
SQUID is biased by a voltage source $V_{b}$ in series with a resistance
$R_{b}=200\,\Omega$. The current $I$ is measured from the voltage
drop across $R_{b}$. An on-chip antenna is used to apply fast flux
pulses to the SQUID. It is represented on the right hand side as an
inductor current-biased with a source $I_{F}.$\textit{ (d)} Four
possible configurations of a one-channel constriction: ground state
(energy $-E_{A}$), Andreev doublet empty; two odd configurations,
with zero energy and definite spin $\pm1/2$, one quasiparticle added
to the contact; last configuration corresponds to spin-singlet double
excitation (energy $+E_{A}$).}

\label{constriction}
\end{figure}
We demonstrate this phenomenon with an experiment on atomic size constrictions,
where the trapping of \emph{a single quasiparticle} is revealed by
the full suppression of the \emph{macroscopic} supercurrent through
its well transmitted conduction channels. We also show that, as anticipated
\cite{Chtchelkatchev}, trapped quasiparticles are long-lived, with
time scales up to hundreds of ${\mu}$s.

We use micro-fabricated mechanically controllable break junctions
\cite{JanQuantro} to obtain aluminum atomic point contacts embedded
in an on-chip circuit, sketched in Fig.\,\ref{constriction}(c)  \cite{supplemental}.
The circuit allows measuring for each atomic contact both the current-voltage
characteristic, from which one determines precisely the ensemble $\left\{ \tau_{i}\right\} $
of the transmissions of its conduction channels \cite{Scheer97},
and its current-phase relation \cite{MLDR}. In order to go reversibly
from voltage to phase biasing, the atomic contact is placed in parallel
with a Josephson tunnel junction (critical current $I_{0}\sim554\,\mathrm{nA}$
much larger than the typical critical current of a one-atom aluminum
contact $\sim50\,\mathrm{nA}$) to form an asymmetric dc SQUID. An
on-chip antenna allows applying fast flux pulses through the SQUID
loop and a superconducting coil is used to apply a dc flux.

In the usual
semi-conductor representation, there is just one pair of Andreev bound
states in a short one-channel constriction, with energies $\pm E_{A}(\delta,\tau)=\pm\Delta\sqrt{1-\tau\sin^{2}(\delta/2)}$
determined by the channel transmission $\tau$ and the phase difference
$\delta$ across it \cite{Chtchelkatchev,supplemental,Michelsen,spin}. In the ground state,
only the Andreev bound state at negative energy is occupied, leading
to a phase dependent term $-E_{A}(\delta,\tau)$ in the total energy,
and a supercurrent $-I_{A}=-\varphi_{0}^{-1}\partial E_{A}/\partial\delta,$
with $\varphi_{0}=\hbar/2e.$ The two Andreev bound states give rise
to the excitation spectrum shown in Fig.\,\ref{constriction}(b),
with a discrete spin-degenerate doublet, localized at the constriction,
at an energy $E_{A}\leq\Delta$ above the ground state. The four lowest-lying
configurations of the system are built from this doublet. Above the
ground state, there are two {}``odd'' configurations (spin 1/2)
with a single excitation of the doublet at $E_{A}$,\emph{ i.e.} with
a quasiparticle trapped in the constriction. In this case the global
energy is zero,\emph{ i.e.} phase independent, and the total supercurrent
is zero \cite{spin}. Finally, there is another spin-singlet configuration
with a double excitation, which carries a supercurrent $+I_{A}$ exactly
opposite to the one in the ground configuration. Hence, the supercurrent
through the constriction is a probe of the configuration of the system.

\begin{figure}
\includegraphics[clip,width=1\columnwidth]{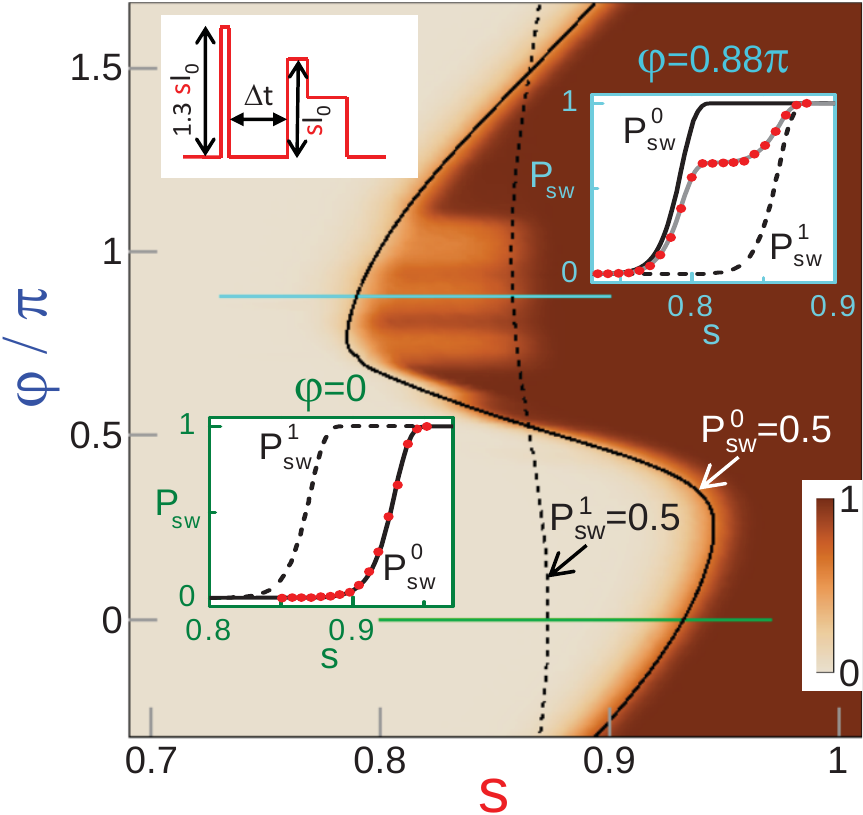}
\caption{(Color online) Color plot of measured switching probability $P_{sw}(s,\varphi)$
for SQUID with contact transmissions \{0.994,0.10,0.10\}.\emph{ Top
left inset:} Measurement protocol. Short prepulses ensure same initial
conditions before each measurement pulse of height $I=sI_{0}$ and
duration $t_{p}=$1\,\textmu{}s (the subsequent lower plateau holds
the voltage to facilitate detection). Delay between prepulse and measurement
pulse is here $\Delta t=2\,\mathrm{\mu s}$. \emph{Main Panel}: Black
curves: theoretical predictions (solutions of $P_{sw}(s,\varphi)\equiv0.5$)
for pristine (solid line) and for poisoned contact (dashed line).
\emph{Insets}: measured $P_{sw}(s)$ (red dots) at fixed flux. \emph{Lower
left}: $\varphi=0$ (green line in main panel). \emph{Upper right}:
$\varphi=0.88\pi$ (cyan line in main panel). In both insets $P_{sw}^{0}(s)\:(P_{sw}^{1}(s))$,
the solid (dashed) line is the theory for the pristine (poisoned)
contact. In the upper right inset, the intermediate line (gray) is
a fit of the data with the linear combination of Eq.~(\ref{eq:CL})
with $p=0.36$.}

\label{AC0}
\end{figure}

In our experiment, the supercurrent through the atomic contact is
accessed through measurements of the {}``switching current'' of
the SQUID, which is the bias current at which the whole device switches
from the supercurrent branch $\left(V=0\right)$ to a finite voltage
state. Because of the large asymmetry, the SQUID switching current
is only slightly modulated around that of the junction by the applied
flux $\phi$. The modulation corresponds essentially to the current-phase
relation of the atomic contact \cite{MLDR}. As the SQUID loop is
small, the phase $\gamma$ across the Josephson junction, the phase
$\delta$ across the atomic contact, and the phase $\varphi=\phi/\varphi_{0}$
related to the external flux $\phi$, are linked \cite{supplemental}
through $\varphi=\delta-\gamma$. To measure the switching current
of the SQUID, current pulses of variable normalized height $s=I/I_{0}$
are applied through the bias line, while monitoring the voltage across
the SQUID. To ensure that the measurements are statistically independent,
additional short prepulses that force switching are applied before
each one of them \cite{supplemental} (top left inset of Fig.\,\ref{AC0}).
The switching probability $P_{sw}\left(s\right)$ is obtained as the
ratio of the number of measured voltage pulses to the number of bias
pulses (typically 10$^{4}$). In Fig.\,\ref{AC0}, we show $P_{sw}(s,\varphi)$
measured at $30\,{\textstyle \mathrm{mK}}$ on one particular SQUID
($\left\{ \tau_{i}\right\} =$\{0.994,0.10,0.10\}). For most flux
values, we observe the generic behavior for Josephson junctions and
SQUIDs, \emph{i.e.} a sharp variation of the probability from 0 to
1 as the pulse height is increased (lower left inset of Fig.\,\ref{AC0}).
However, in a broad flux range $0.7\pi<\varphi<1.1\pi$ around $\pi$,
the behavior is completely unusual: $P_{sw}(s)$ increases in two
steps and displays an intermediate plateau (top right inset of Fig.\,\ref{AC0}). 

Precise comparison between experiment and theory is performed using
an extension \cite{MLDR,supplemental} of the well-known model describing
the switching of a Josephson junction as the thermal escape of a particle
over a potential barrier \cite{escape th}. For our
SQUIDs, the potential is dominated by the Josephson energy of the
junction but contains a small contribution $\sum_{i}c_{i}E_{A}(\gamma+\varphi,\tau_{i})$
from the atomic contact, which depends on $\left\{ \tau_{i}\right\} $
and on the configuration of its Andreev levels ($c_{i}=-1$ if channel
$i$ is in its ground state, $c_{i}=1$ in excited singlet, and $c_{i}=0$
in an odd configuration).%
The predictions for $P_{sw}(s)$ are shown as lines in the insets
of Fig.\,\ref{AC0}. Whereas the data taken at $\varphi=0$ are well
fitted by theory with all channels in the ground state, those taken
at $\varphi=0.88\pi$ are not. However, they can be very precisely
accounted for by the weighted sum of the theoretical curves $P_{sw}^{0}(s)$
and $P_{sw}^{1}(s)$ corresponding, respectively, to the {}``pristine
contact'' (\emph{i.e.} with all channels in their ground configuration),
and to the {}``poisoned'' contact (\emph{i.e.} with its most transmitted
channel in an odd configuration). This is the case in the whole flux
region where the measured switching curves have an intermediate plateau,
showing that \begin{equation}
P_{sw}(s,\varphi)=(1-p\left(\varphi\right))P_{sw}^{0}(s,\varphi)+p\left(\varphi\right)P_{sw}^{1}\left(s,\varphi\right).\label{eq:CL}\end{equation}
The function $1-p\left(\varphi\right)$ describes the height of the
intermediate plateau in $P_{sw}(s)$. A similar analysis was performed
for other SQUIDs formed with atomic contacts having one highly transmitted
channel \cite{supplemental} and in all cases, $P_{sw}(s)$ shows
a plateau delimited by the predictions for the pristine and the poisoned
contact in a broad phase range around $\pi$. The fact that the data
are precisely accounted for by this linear combination induces us
to interpret the coefficient $p$ as the poisoning probability, \emph{i.e.}
the probability for the atomic contact to have a quasiparticle trapped
in its most transmitting channel.

\begin{figure}
\begin{centering}
\includegraphics[clip,width=1\columnwidth]{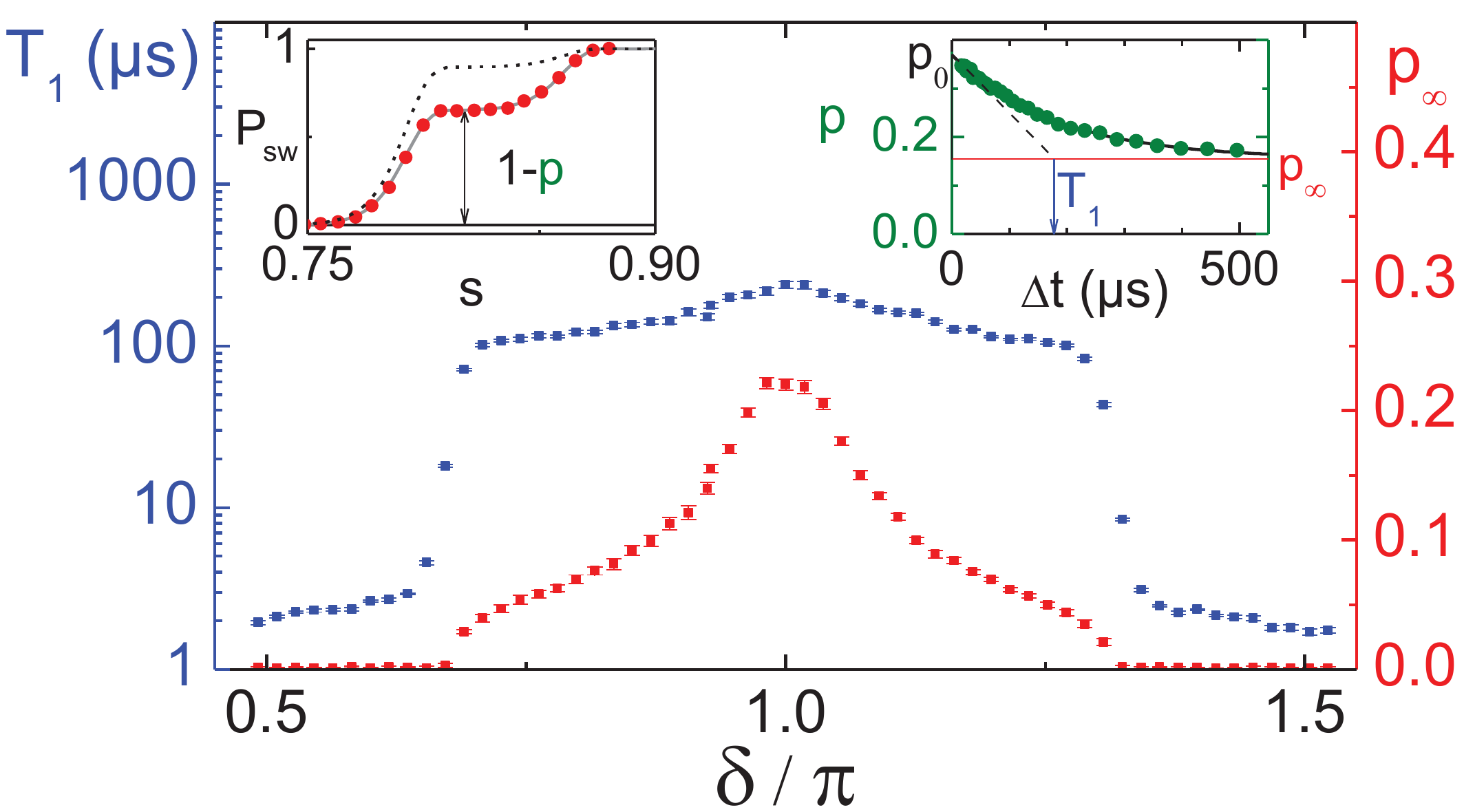}
\par\end{centering}

\caption{(Color online) Relaxation time $T_{1}$ (blue points) and asymptotic
poisoning probability $p_{\infty}$(red points) as a function of the
phase $\mathrm{\delta}$ imposed for a time $\Delta t$ between prepulse
and measurement pulse. Data taken on SQUID with atomic contact \{0.994,0.10,0.10\}
at $30\,\mathrm{mK}$. Measurements at $\left|\delta-\pi\right|>0.5\pi$
show very fast relaxation that could not be resolved reliably in our
setup. \emph{Left inset}: $1-p$ is the height of the intermediate
plateau in $P_{sw}(s)$. Dashed line: $P_{sw}(s)$\emph{ }found\emph{
}for $\Delta t\gg100\mathrm{\text{\textmu}s.}$ \emph{Right inset}:
Typical time evolution of $p$, from where $T_{1}$ and $p_{\infty}$
are extracted.}

\label{pollution1} 
\end{figure}

We have found that at fixed $s$ and $\varphi$, the poisoning probability
$p$ depends exponentially on the delay $\Delta t$ between the prepulse
and the measurement pulse (Fig.~3 right inset): a fit of the form
$p(\Delta t)=p_{\infty}+(p_{0}-p_{\infty})\exp(-\Delta t/T_{1})$
gives the initial poisoning just after the prepulse $p_{0}$, the
asymptotic value at long times $p_{\infty}$, and the relaxation time
$T_{1}$. To obtain a meaningful measurement of the phase dependence
of the relaxation, we had to implement a refined protocol \cite{supplemental}
involving flux pulses applied through the fast flux line within the
time interval $\Delta t$. It allows probing the relaxation from a
fixed $p_{0}$, with measurement occuring always at the same flux,
the only adjustable parameter being the phase $\delta$ during the
waiting time. Both $p_{\infty}\left(\delta\right)$ and $T_{1}\left(\delta\right)$,
measured at 30 mK, are shown in Fig.\,\ref{pollution1} for the same
SQUID as in Fig.\,\ref{AC0}. Their phase dependence is symmetric
and peaked at $\delta=\pi,$ where the relaxation is the slowest and
$p_{\infty}$ the largest. A rapid decay of $T_{1}$ by almost two
orders of magnitude and a drop of $p_{\infty}$ to 0 are observed
at $\left|\delta-\pi\right|\simeq0.3\pi.$ The overall shape of both
$p_{\infty}\left(\delta\right)$ and $T_{1}\left(\delta\right)$ remains
very similar when temperature is varied \cite{supplemental}. The
relaxation time $T_{1}$ falls rapidly with temperature, and becomes
too short to be measured above 250\,mK.%
Similar data \cite{supplemental} taken on a variety of atomic contacts
show that the phase interval in which poisoning occurs reduces when
the transmission of the most transmitting channel diminishes. For
channels with all transmissions smaller than 0.7, the poisoning probability
$p$ was too small to be measured. An important observation is that
when two switching prepulses are applied (instead of a single one)
with more than $1\,\text{\textmu s}$ delay between them, the first
one has no effect, which indicates that quasiparticles created by
switching matter only during 1~\textmu{}s. This experimental observation,
plus the fact that diffusion is expected to efficiently drain away
from the constriction the quasiparticles created by the prepulse \cite{supplemental},
allow us to conclude that the residual quasiparticle density is constant
during the poisoning probability relaxation, and that it does not
originate from the switching pulses (contrary to $p_{0}$). The fact
that $p_{\infty}\neq0$ proves that quasiparticles are present in
the continuum in the steady state, as found in other experiments \cite{Shaw,Martinis,de Visser}.

\begin{figure}[h]
\begin{centering}
\includegraphics[clip,width=1\columnwidth]{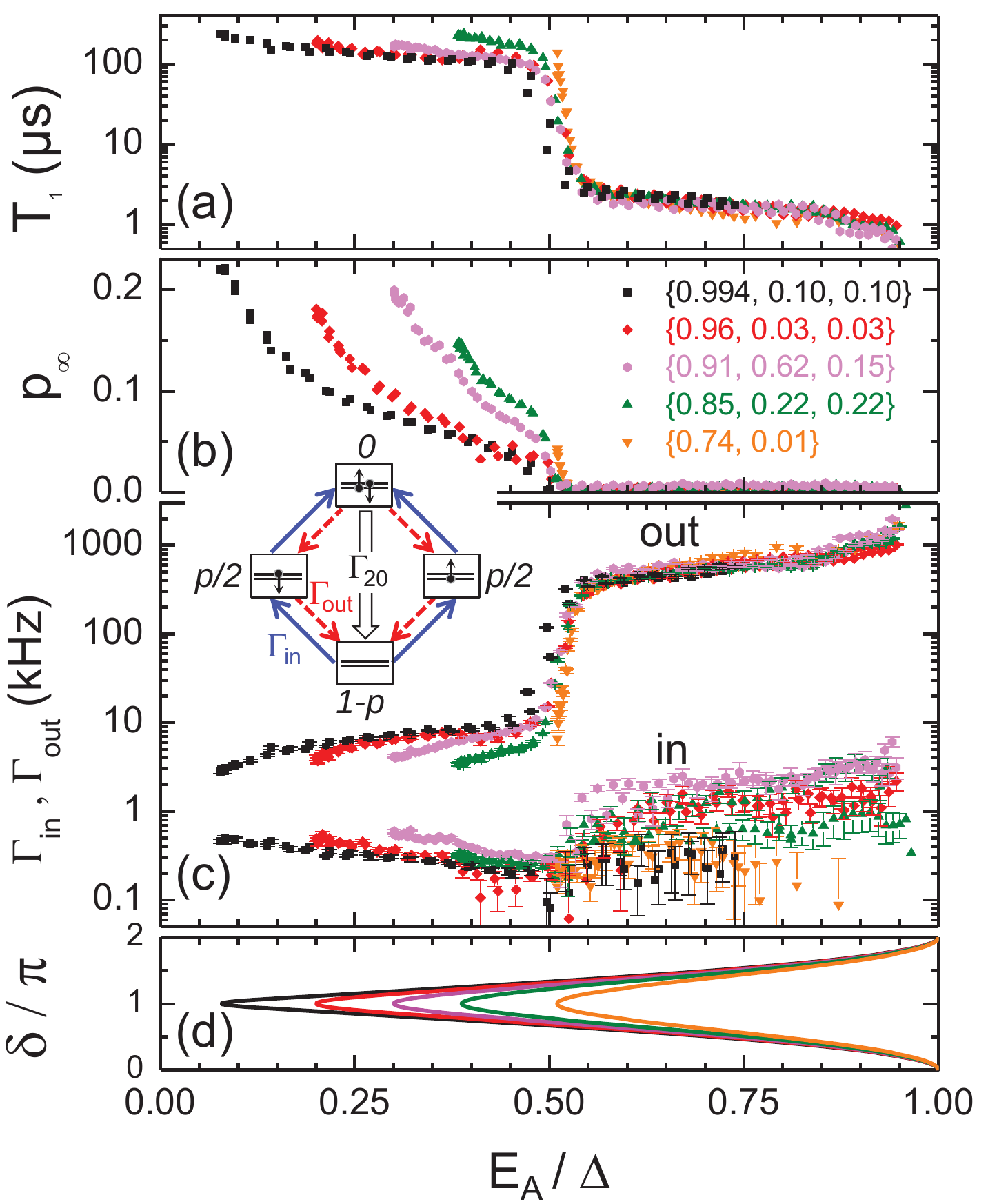} 
\par\end{centering}

\caption{(Color online) Relaxation data for five different atomic contacts
(transmissions are given in panel (b)): (a) relaxation time $T_{1},$
(b) asymptotic poisoning probability $p_{\infty},$ (c) rates $\Gamma_{\mathrm{in}}$
and $\Gamma_{\mathrm{out}}$ as a function of normalized Andreev state
energy $E_{A}/\Delta$ of most transmitting channel. There is a sharp
threshold at $E_{A}/\Delta\approx0.5$ for all contacts. Minimal value
of $E_{A}/\Delta$ is $\sqrt{1-\tau}$ (0.08 for the black points,
0.5 for the orange points) and is reached when $\delta=\pi,$ as shown
in (d). Inset of panel (c): Rates $\Gamma_{\mathrm{in}}$ (resp. $\Gamma_{\mathrm{out}}$)
are for processes increasing (resp. decreasing) the number of quasiparticles
in contact. The relaxation rate from the excited singlet to the ground
state is assumed to be much faster than all other rates. The occupation
of all four configurations is given in italic letters: $1-p$ for
the ground state, $p/2$ for each of the odd configurations, and $0$
for the excited state.}

\label{Gammas} 
\end{figure}

When quasiparticles jump between the Andreev states and the states
in the continuum, transitions arise among the four configurations
accessible to the Andreev doublet, as shown in the inset of Fig.~\ref{Gammas}.
All the microscopic processes involved are in principle rather slow
because they either require energy absorption or the presence of quasiparticles
in the continuum \cite{Chtchelkatchev}. We define the rates $\Gamma_{\mathrm{in}}$
and $\Gamma_{\mathrm{out}}$ corresponding to an increase or a decrease
of the number of quasiparticles in the contact. Because we see no trace of the state with a double excitation neither in these data, nor in preliminary spectroscopic measurements, we assume
that the relaxation rate $\Gamma_{20}$ to the ground state is much
larger than $\Gamma_{\mathrm{in}}$ and $\Gamma_{\mathrm{out}}.$
From a simple master equation \cite{supplemental} for the population
$1-p$ of the ground state, and $p/2$ of each of the odd configurations,
one gets\begin{equation}
T_{1}=\frac{1}{\Gamma_{\mathrm{out}}+3\Gamma_{\mathrm{in}}},\hspace{20pt}p_{\infty}=\frac{2\Gamma_{\mathrm{in}}}{\Gamma_{\mathrm{out}}+3\Gamma_{\mathrm{in}}}\label{eq:TR}\end{equation}
and the flux dependence of $\Gamma_{\mathrm{in}}$ and $\Gamma_{\mathrm{out}}$
can then be extracted from the data. Then, instead of plotting the
results as a function of the applied flux $\delta$ like in Fig.~3,
we choose for the x-axis the Andreev energy $E_{A}\left(\delta\right)$
of the most transmitting channel. The results for five contacts are
shown in Fig.~\ref{Gammas}, together with the relaxation time $T_{1}$
and asymptotic poisoning probability $p_{\infty}.$%
Although the Andreev energy is clearly not the only relevant parameter,
the rates for all contacts roughly coincide. The most apparent differences
are in the asymptotic poisoning probability $p_{\infty}$ which, for
a given $E_{A},$ diminishes when the transmission of the most transmitting
channel increases. Two distinct regimes are evidenced in Fig.~\ref{Gammas}:
when $E_{A}/\Delta>0.5,$ the relaxation time is very short and the
asymptotic poisoning is negligible. In terms of rates, $\Gamma_{\mathrm{in}}$
is smaller than $\Gamma_{\mathrm{out}}$ by 2 to 3 orders of magnitude.
In contrast, when the Andreev energy lies deep in the gap $\left(E_{A}/\Delta<0.5\right)$,
relaxation is much slower and the asymptotic poisoning probability
becomes sizable. This regime corresponds to a smaller ratio $\Gamma_{\mathrm{out}}/\Gamma_{\mathrm{in}}$.
The sharp threshold at $E_{A}/\Delta\simeq0.5,$ where $\Gamma_{\mathrm{out}}$
drops by two orders of magnitude, is observed for all the measured contacts
with highest transmission above 0.74.
Furthermore, no poisoning was observed in contacts in which all channels
had transmissions below 0.74, the Andreev state energy then being
always larger than $\Delta\sqrt{1-0.74}\sim0.5\,\Delta.$ In contrast,
we have found that in contacts with more than one highly transmitting
channel, poisoning can affect several channels at once \cite{supplemental}.
Presently, we do not have an explanation for the energy dependence
of the rates: the mechanisms commonly used to describe quasiparticle
dynamics in superconductors (recombination, phonon emission and absorption)
do not lead to such a sharp threshold at energy $0.5\,\Delta$. 

Let us mention that it is possible to untrap quasiparticles. For example,
we have implemented an efficient {}``antidote'' protocol \cite{supplemental}
based on dc flux pulses that bring the Andreev states at the gap edge
from where the trapped quasiparticle can diffuse away into the electrodes.
Furthermore, it is possible to avoid altogether poisoning: when the
large scale on-chip wires connecting to the SQUID are made out of
either a normal metal \cite{MLDR} or a superconductor with a lower
gap than the device \cite{supplemental}, they act as good quasiparticle
traps and poisoning is never observed.

To conclude, we have performed the first observation and characterization
of single quasiparticles trapped in Andreev bound states. The long
lifetimes that we have measured open the way to individual spin manipulation
and to superconducting spin qubits \cite{Chtchelkatchev}. Moreover
the complete suppression of the macroscopic supercurrent when a single
quasiparticle is trapped shows that a superconducting quantum point
contact can be seen as a very efficient quasiparticle detector. Finally,
let us mention that quasiparticle trapping, which is likely to be
a generic phenomenon in superconducting weak links, could be detrimental
in some situations. It could be the case for experiments proposed
to detect {}``Majorana bound states'' in condensed matter systems
\cite{Fu} since their topological protection relies on parity
conservation. 

We thank H. Grabert, A. Levy Yeyati, J. Martinis, V. Shumeiko and
D. Urban for enlightening discussions. We gratefully acknowledge help
from other members of the Quantronics group, in particular P. Senat
and P.F. Orfila for invaluable technical assistance. Work partially
funded by ANR through projects CHENANOM, DOC-FLUC and MASQUELSPEC, and by C'Nano IdF.

\end{document}


\title{Supplemental material for: {}``Evidence for long-lived quasiparticles
trapped in superconducting point contacts''}

\author{M.\ Zgirski}
\author{L.\ Bretheau}
\author{Q.\ Le Masne}
\author{H.\ Pothier}
\email[Corresponding author~: ]{hugues.pothier@cea.fr}
\author{D.\ Esteve}
\author{C.\ Urbina}

\affiliation{Quantronics Group, Service de Physique de l'État Condensé (CNRS,
URA\ 2464), IRAMIS, CEA-Saclay, 91191 Gif-sur-Yvette, France}

\begin{abstract}
We first discuss the various representations of the spectrum of Andreev
bound states, then the sample design and fabrication, and the switching
current measurements. The details of the poisoning dynamics measurements
are then reported, followed by a description of a procedure, nicknamed
{}``poisoning antidote'', to untrap quasiparticles from the point
contact. A discussion of the various processes at play follows, with
a presentation of the results in terms of poisoning and unpoisoning
rates as a function of the energy of the Andreev states. Details on
the dynamics of our circuit are discussed in appendix A. Appendix
B focuses on the calculation of the relaxation rate from the spin-singlet
double excitation.
\end{abstract}
\maketitle

\section{Andreev bound states representations}

\begin{figure}[h]
\begin{centering}
\includegraphics[width=0.7\columnwidth,clip]{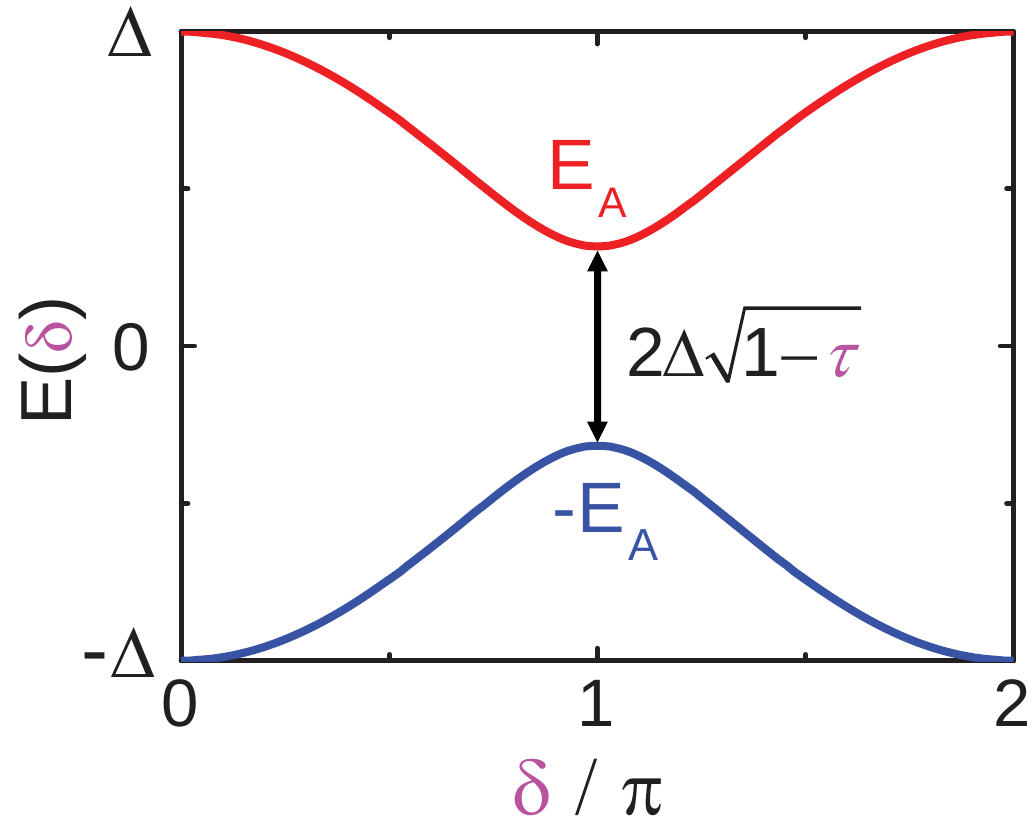} 
\par\end{centering}

\caption{Phase $(\delta$) dependence of the Andreev bound states energies
$\pm E_{A}$ in a short transport channel of transmission $\tau.$
The energy $\Delta$ is the superconducting gap.}

\label{ABS} 
\end{figure}

The widely used picture for Andreev bound states (ABS) in a short
single channel conductor is the semiconductor representation \cite{Tinkham,Bagwell}
shown in Fig.\,\ref{ABS}, assuming spin degeneracy (continuum states
at energies smaller than $-\Delta$ and larger than $\Delta$ are
not shown). The lines correspond to the energies of the available
energy levels. In the ground state, all levels at negative energies
are occupied. The total energy has a single phase-dependent term $-E_{A}(\delta)$,
arising from the lowest ABS. When adding a single quasiparticle to
the system (quasihole or quasielectron), one accesses the two odd
configurations, which correspond to the ABS being both empty or both
occupied. Both configurations have zero energy ($-E_{A}+E_{A}$ or
$0+0$), in absence of magnetic field or spin-orbit coupling \cite{Michelsen,Chtchelkatchev}.
Finally, adding two quasiparticles or exciting the system with photons
at energy $2E_{A}$ gives access to the spin-singlet double-excitation
configuration, represented here with the lowest ABS being empty and
the top one being occupied. The excitation spectrum shown in Fig.~1d
of the Letter directly follows: at a given phase $\delta,$ two possible
single particle excitations have an energy $E_{A}$ above the ground
state energy; the cost of the double excitation is $2E_{A}.$

\begin{figure}[h]
\begin{centering}
\includegraphics[clip,width=0.8\columnwidth]{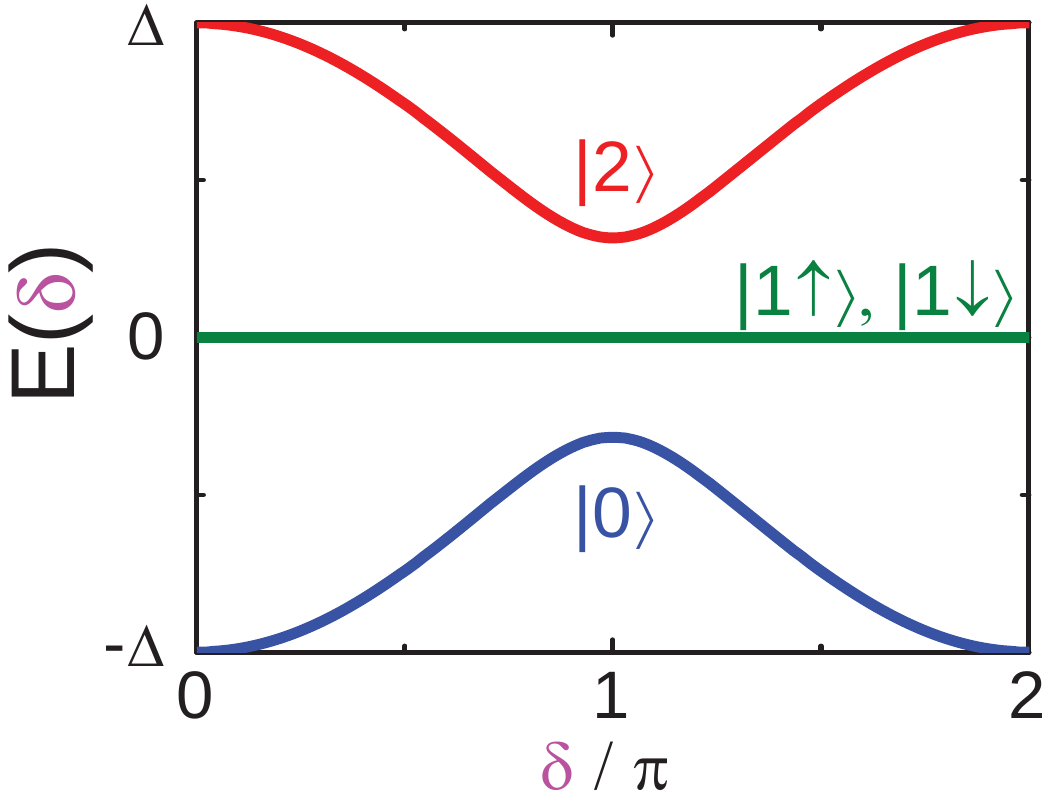} 
\par\end{centering}

\caption{Energy of the different configurations accessible to a single channel.
The energy of the ground configuration, labeled $\left|0\right\rangle $,
is that of the lowest Andreev level $-E_{A}$ of Fig.~\ref{ABS}.
Two odd configurations, labeled $\left|1\uparrow\right\rangle $ and
$\left|1\downarrow\right\rangle $, with one quasiparticle added to
the ground state, have zero energy and a definite spin\cite{Chtchelkatchev}.
In the configuration $\left|2\right\rangle $, only the excited Andreev
state $+E_{A}$ is occupied.}

\label{levels2} 
\end{figure}

An alternative representation, used in Ref.~~\onlinecite{Chtchelkatchev}
shows the energy of the various configurations as a function of the
phase difference. Without spin-dependent interactions, one obtains
Fig.~\ref{levels2}. %

As discussed in the main paper, the vanishing of the supercurrent
is interpreted as arising from poisoning of the Andreev doublet by
a single quasiparticle, \textit{i.e.} a transition from the ground
state$\left|0\right\rangle $ to an odd configuration $\left|1\uparrow\right\rangle $
or $\left|1\downarrow\right\rangle $. We think that the following
description, although not rigorous, helps clarifying the nature of
the Andreev states and their spin structure: in the ground state of
the system, a Cooper pair state (spin singlet state) is formed at
the contact with an energy $-E_{A}\left(\delta\right)\in\left[-\triangle,0\right]$,
through Andreev reflections in the superconductors on both sides.
Mimicking the BCS theory in second quantization formalism\cite{Tinkham},
and inspired by Ref.~{[}\onlinecite{Vecino,Meng}{]}, the superconducting
ground state wave-function at the contact can be thought of as \begin{equation}
\left|\Psi_{0}\right\rangle =\left|\Psi_{cont}\right\rangle \otimes\left(u_{A}+v_{A}c_{A\uparrow}^{\dagger}c_{A\downarrow}^{\dagger}\right)\left|0\right\rangle ,\end{equation}
with $\left|\Psi_{cont}\right\rangle $ describing the states with
energy lying below $-\Delta$ (continuum states) and $\left|0\right\rangle $
is the vacuum. We have singled-out from the BCS wave-function the
state {}``A'' corresponding to the lowest energy Andreev state.
The creation operators $c_{A\uparrow,\downarrow}^{\dagger}$ create
an electron at the Fermi level in the channel, with either spin. Hence,
the coefficients $u_{A}$ and $v_{A}$ differ only by a phase, and
\begin{equation}
\left|\Psi_{0}\right\rangle =\left|\Psi_{cont}\right\rangle \otimes\frac{1}{\sqrt{2}}\left(1+e^{i\delta_{A}}c_{A\uparrow}^{\dagger}c_{A\downarrow}^{\dagger}\right)\left|0\right\rangle .\end{equation}
Thus, the Andreev energy $-E_{A}$ can be seen as the condensation
energy of the {}``Andreev Cooper pair''.

The addition of one quasiparticle to the system at the lowest possible
energy is described by one of the two Bogoliubov operators $\gamma_{A\uparrow}^{\dagger}=u_{A}^{*}c_{A\uparrow}^{\dagger}-v_{A}^{*}c_{A\downarrow}$
or $\gamma_{A\downarrow}^{\dagger}=u_{A}^{*}c_{A\downarrow}^{\dagger}+v_{A}^{*}c_{A\uparrow}$,
which lead to the {}``odd'' states \begin{equation}
\left|\Psi_{1\uparrow}\right\rangle =\gamma_{A\uparrow}^{\dagger}\left|\Psi_{0}\right\rangle =\left|\Psi_{cont}\right\rangle \otimes c_{A\uparrow}^{\dagger}\left|0\right\rangle \end{equation}
 and \begin{equation}
\left|\Psi_{1\downarrow}\right\rangle =\gamma_{A\downarrow}^{\dagger}\left|\Psi_{0}\right\rangle =\left|\Psi_{cont}\right\rangle \otimes c_{A\downarrow}^{\dagger}\left|0\right\rangle ,\end{equation}
both having a single electron occupying with certainty the state at
the Fermi energy. The energy of the odd states is higher than the
ground state energy by the condensation energy $E_{A}$ of the Cooper
pair in the Andreev bound state. For completion, the upper Andreev
state $\left|\Psi_{2}\right\rangle $ is obtained when two quasiparticles
are added to the ground state, at an energy cost $2E_{A}:$ \begin{multline}
\begin{alignedat}{1}\left|\Psi_{2}\right\rangle  & =\gamma_{A\downarrow}^{\dagger}\gamma_{A\uparrow}^{\dagger}\left|\Psi_{0}\right\rangle =-\gamma_{A\uparrow}^{\dagger}\gamma_{A\downarrow}^{\dagger}\left|\Psi_{0}\right\rangle \\
 & =\left|\Psi_{cont}\right\rangle \otimes\left(u_{A}^{*}c_{A\uparrow}^{\dagger}c_{A\downarrow}^{\dagger}-v_{A}^{*}\right)\left|0\right\rangle .\end{alignedat}
\end{multline}
Apart from a non-physical global phase, this state can be rewritten
as $\left|\Psi_{cont}\right\rangle \otimes\frac{1}{\sqrt{2}}\left(-1+e^{i\delta_{A}}c_{A\uparrow}^{\dagger}c_{A\downarrow}^{\dagger}\right)\left|0\right\rangle ,$
which resembles the ground state $\left|\Psi_{0}\right\rangle $:
it contains an excited Cooper pair (excited singlet). As a matter
of fact, the present experiment was initially designed to perform
the spectroscopy of the transition between $\left|\Psi_{0}\right\rangle $
and $\left|\Psi_{2}\right\rangle ,$ a goal that we haven't reached
yet.

\section{Sample fabrication and measurement setup}

\begin{figure*}
\begin{centering}
\includegraphics[clip,width=2\columnwidth]{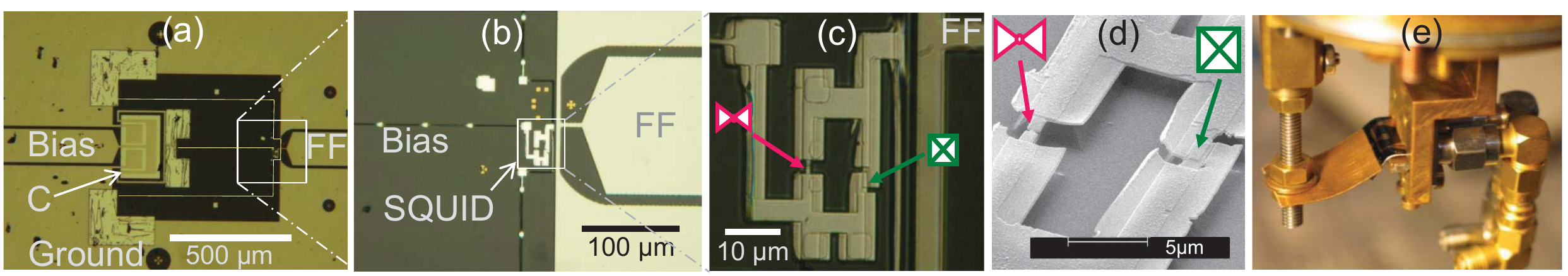} 
\par\end{centering}

\caption{(a) Large scale photograph of the sample, showing the end of the fast
flux (FF) and current bias coplanar lines. The bias line and the ground
plane form planar capacitors with a common, floating and superconducting
rectangular electrode placed below, implementing the capacitor $C$.
From there the connections to the SQUID are made with 0.9~\textmu{}m-wide
Al wires which are not visible at this scale, and have been redrawn
for clarity. They appear on the photograph (b) as three dark lines
with regularly spaced bright spots, one connected to the bias line,
the two others to the ground plane. The bright spots are wider pillars
that hold on the substrate since they are large enough not to be completely
freed during the etching step. The fast flux coplanar line ends with
a short close to the SQUID, the current in the upper half of the short
creating a magnetic flux in the loop. Photograph (c) shows the SQUID
loop, with the Josephson junction on the right and the suspended bridge
where the atomic contacts are formed on the left. Five small square
gold electrodes, intended as quasiparticle traps \cite{QPtraps},
are barely visible through the Aluminum layers. (d) SEM micrograph
of the SQUID, seen under an angle: two angle evaporation of aluminum
define the superconducting loop with a tunnel junction on the right
arm, and a suspended micro-bridge on the left arm, which is broken
at low temperature to form atomic contacts. (e) View of the sample
holder, with a bent sample. The sample is clamped between the launchers
of two SMA connectors, which are visible on the right-hand side, and
a small metallic plate hold with two screws at the bottom. The sample
is bent with a brass blade held by a rod moving vertically (on the
left). For this photograph, the bending was exagerated, and the superconducting
coil placed immediately above the sample was removed.}

\label{setup-2-1} 
\end{figure*}

\begin{figure*}
\begin{centering}
\includegraphics[clip,width=1\columnwidth]{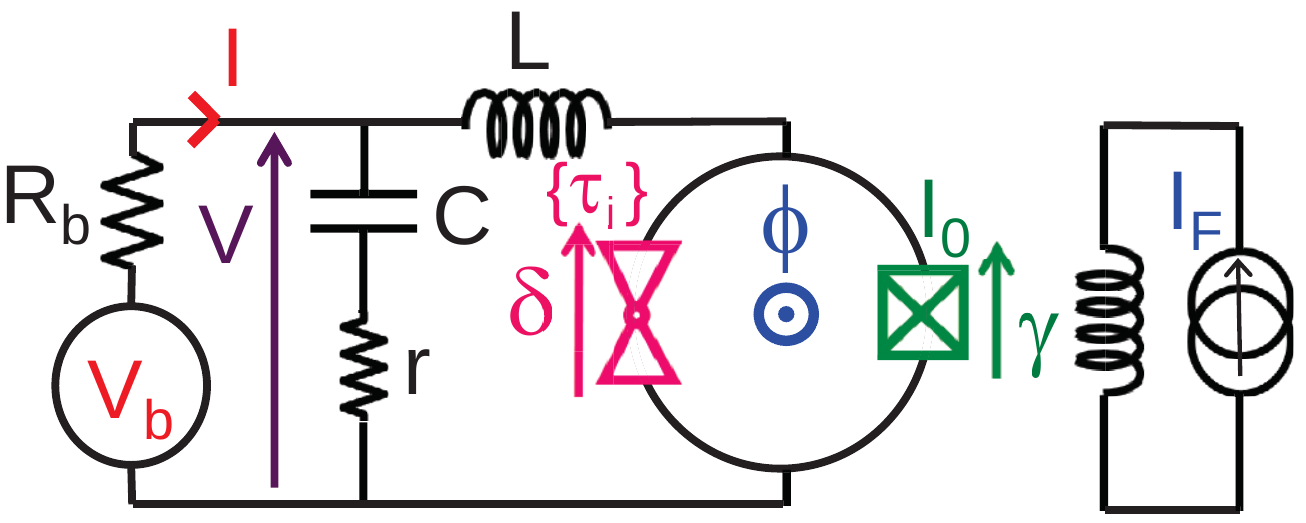} 
\par\end{centering}

\caption{Full schematic of the sample. The SQUID is biased through an on-chip
$LC$ circuit ($L\simeq1.6\,\mathrm{nH},$ $C\simeq65\,\mathrm{pF}$).
The losses in the capacitor are modelled by series resistance $r\simeq0.5\,\Omega.$
The voltage $V$ measured across the full on-chip circuit corresponds,
at low frequency, to the voltage drop across the SQUID. }

\label{FullSetup} 
\end{figure*}

The samples were fabricated on a polished, 500~\textmu{}m-thick Kapton
substrate, which is insulating and elastic. A suspended bridge is
fabricated by e-beam lithography of a thin constriction in an aluminum
film, and etching of 1~\textmu{}m of an underlying polyimide layer.
Optical micrographs of the sample at various scales, a SEM micrograph
of the core of the sample and a photograph of the break-junction mount
used to obtain atomic contacts are shown in Fig.\,\ref{setup-2-1}.
The bridge is placed in a superconducting loop containing also a 2.8~\textmu{}m$^{\text{2}}$
Josephson junction fabricated at the same time by shadow evaporation
(70~nm Al, oxidation in an $\mathrm{Ar-O_{2}}$ (85\%-15\%) mixture
at 18~mbar for 5~min, 54~nm Al), hence forming a SQUID\cite{TheseChauvin,MLDR}.
The SQUID is connected with thin (124~nm), narrow (0.9~\textmu{}m)
and long (0.4 and 0.8~mm) Al wires corresponding to a total inductance
$L$ in the nH range. At a larger scale, the connecting wires overlap
over 0.01\,mm$^{\text{2}}$ a floating underlying 30~nm-thick aluminum
electrode, which was covered with five heavily oxidized 1.5~nm-thick
aluminum layers, forming a capacitor $C\simeq65\,\mathrm{pF}$, hence
shunting the connecting lines at high frequency. The resonance frequency
$(2\pi\sqrt{LC})^{-1}\simeq0.5\,\mathrm{GHz}$ and the losses of the
corresponding tank circuit (see Fig.~\ref{FullSetup}), modelled
by series resistance $r\simeq0.5\,\Omega,$ were measured by microwave
reflectometry in a separate experiment with the same sample. From
the capacitor, a 50\,$\Omega$ coplanar waveguide makes the connection
towards a mm-size connection pad placed at the edge of the sample. 

The sample was cooled in a dilution refrigerator operated down to
25\,mK. The substrate is clamped at one of its edges, between a small
copper plate placed underneath and the half-cylindrical central pins
of two SMA launchers, which connect to the bias and fast flux lines\cite{These QLM}
(see Fig.~\ref{setup-2-1}(e)). Connections of the ground plane to
the sample holder are achieved in the same manner\cite{Indium}. The
atomic contacts are obtained by bending the sample with a pusher placed
on the side opposite to the contact pads. The contacts formed with
this setup are very stable. For example, the contact discussed in
the Letter, labeled AC1 in the following, was kept unchanged during
2 months, till we decided to form another one. A shielded superconducting
coil (which had been withdrawn before taking the picture in Fig.~\ref{setup-2-1}(e))
is placed a mm above the SQUID, for dc flux biasing. Bias lines connected
to the SQUID and to the antenna are coaxial lines heavily attenuated
at various stages of the cryostat, with total attenuation of 55 and
30dB, respectively, in order to damp the current noise of the components
placed at higher temperature. The current and voltage measurement
lines are twisted pairs, equipped with micro-fabricated microwave
filters\cite{filtersHlS}.%

\section{determination of the transmission coefficients}

\begin{figure}[h]
\begin{centering}
\includegraphics[clip,width=1\columnwidth]{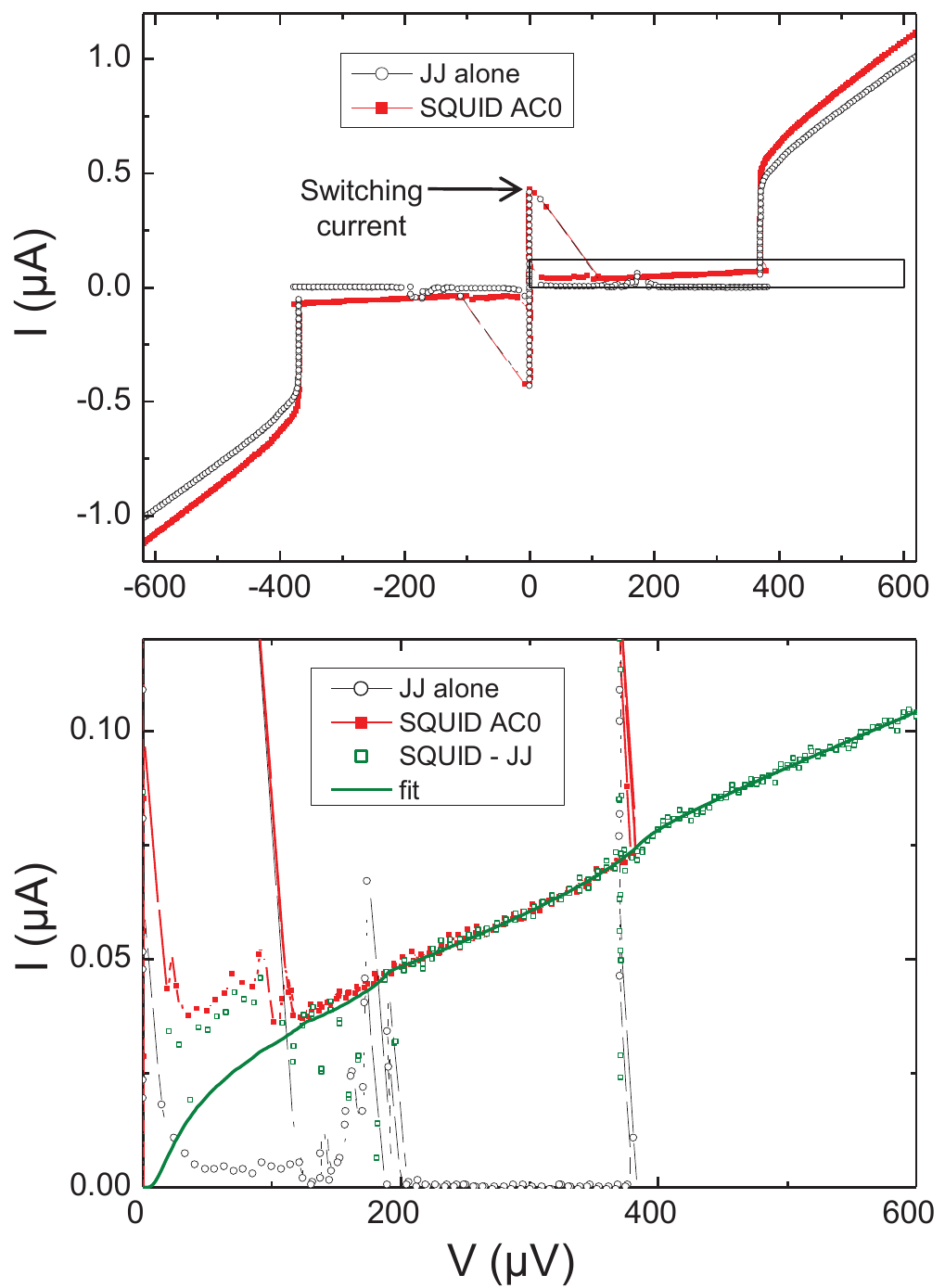} 
\par\end{centering}

\caption{Top: Current-voltage $\left(I-V\right)$ characteristics of the Josephson
junction (black open symbols) and of a SQUID with an atomic contact
labeled AC0 (red filled symbols). The switching current of the SQUID,
indicated with an arrow, is modulated by the applied flux. Bottom:
same data, and (green open symbols) $I-V$ characteristic of the atomic
contact alone obtained by subtraction of the one of the junction from
that of the SQUID. The transmissions \{0.95,0.45,0.10\} are found
by fitting this $I-V$ characteristic with the theory of Multiple
Andreev Reflexions (green solid curve), excluding the region $\left|V\right|<200\text{\,\textmu}$eV
where the $I-V$ characteristic of the Josephson junction presents
resonances.}

\label{IV} 
\end{figure}

The current-voltage characteristic of the Josephson junction alone,
taken after opening completely the atomic contact, is shown with black
open symbols in Fig.\,\ref{IV}. As observed in other experiments
with similar junctions\cite{MLDR,These QLM,Greibe}, some current
is found at sub-gap voltages, here for $\left|V\right|<200\,{\mu}\mathrm{eV}$.
The large scale resistance is $R_{JJ}=550\,\Omega$. The same figure
presents with red solid symbols the $I-V$ characteristic of a SQUID
with an atomic contact: as compared to the previous curve, the large
scale conductance is slightly increased, and a significant sub-gap
current is visible. The difference between the two characteristics,
which represents the contribution of the atomic contact to the dissipative
current, is shown in the inset with green open symbols. It is fitted
with the theory\cite{AverinBardas,Bratus95,Cuevas96} of Multiple
Andreev Reflections (MAR)\cite{Klapwijk1982} in order to obtain the
transmission of its conduction channels\cite{Scheer97}. The region
$\left|V\right|<200\,{\mu}\mathrm{eV}$, where sub-gap current was
already found in the $I-V$ characteristic of the Josephson junction
by itself, was excluded from the fit. As a consequence, the accuracy
on the determination of the transmissions of the channels is not as
good as in experiments with atomic contacts alone\cite{ReviewALR}.
It is here of the order of 1\% for the largest transmission, and 3\%
for the second largest. As an example, the contact corresponding to
Fig.\,\ref{IV}, called AC0 in the following, was found to have 3
channels, as most one-atom Aluminum contacts do; their transmissions,
deduced from the fit of the $I-V$ characteristic, are \{0.95,0.45,0.10\}.
The corresponding fit is shown in the inset of Fig.~\ref{IV} as
a solid line.%

\section{Switching current measurements\label{sec:Switching-current-measurements}}

The supercurrent associated with the phase-dependence of the Andreev
bound states energy is accessed through measurements of the {}``switching
current'' of the SQUID (see Fig.~\ref{IV}). The geometrical inductance
of the loop, of the order of 20~pH\cite{These QLM}, can be neglected
when compared to the Josephson inductances of the junction ($L_{JJ}=\varphi_{0}/I_{0}\simeq0.6\,\mathrm{nH}$)
and of the atomic contact ($L_{AC}\sim4\varphi_{0}^{2}/\tau\triangle\simeq10\,\mathrm{nH}$).
Therefore, the phase $\gamma$ across the Josephson junction, the
phase $\delta$ across the atomic contact and the flux phase $\varphi=2\pi\phi/\phi_{0}$
are related through $\varphi=\delta-\gamma$. 

The modulation of the switching current of the SQUID can be qualitatively
understood as follows\cite{These QLM}: when a flux is applied, the
flux phase $\varphi$ drops mainly across the weakest link in the
loop, \emph{i.e.} the atomic contact. In contrast, when a bias current
$I$ is applied, it flows essentially in the arm with the largest
critical current, which is the one with the Josephson junction. As
a consequence, $\gamma\simeq\gamma(s)\equiv\mathrm{arcsin}\left(s\right)$
and $\delta\approx\varphi+\gamma(s),$ with $s=I/I_{0}$ the normalized
current and $I_{0}$ the critical current of the Josephson junction
alone. The current circulating in the SQUID loop, which is due to
the phase-biased atomic contact, is therefore $I_{at.c.}\left(\varphi+\gamma(s)\right)=\sum_{i}c_{i}I_{A}(\varphi+\gamma(s),\tau_{i})$,
and the total current through the Josephson junction $I-I_{at.c.}\left(\varphi+\gamma(s)\right)$.
Since $I_{A}\ll I_{0},$ the switching current of the SQUID will be
close to the switching current $I_{sw}^{JJ}=s_{sw}^{JJ}I_{0}$ of
the junction alone, and $\gamma(s)\simeq\gamma\left(s_{sw}^{JJ}\right).$
In the experiments presented here, switching for the junction alone
occurs around $s_{sw}^{JJ}\approx0.88,$ and $\gamma\left(s_{sw}^{JJ}\right)\approx0.3\pi.$
As a consequence, switching for the SQUID occurs when $I-I_{at.c.}\left(\varphi+\gamma(s_{sw}^{JJ})\right)\approx I_{sw}^{JJ}$:
the average switching current is that of the junction alone, and the
modulation corresponds, in a first approximation, to the current-phase
relation of the atomic contact, with a phase offset. The modulation
of the switching current by the flux is therefore a direct measurement
of the current-phase relation of the atomic contact and not, as one
could believe at first sight, a measure of the \textit{switching}
current of the atomic contact alone.

\begin{figure}[h]
\begin{centering}
\includegraphics[clip,width=1\columnwidth]{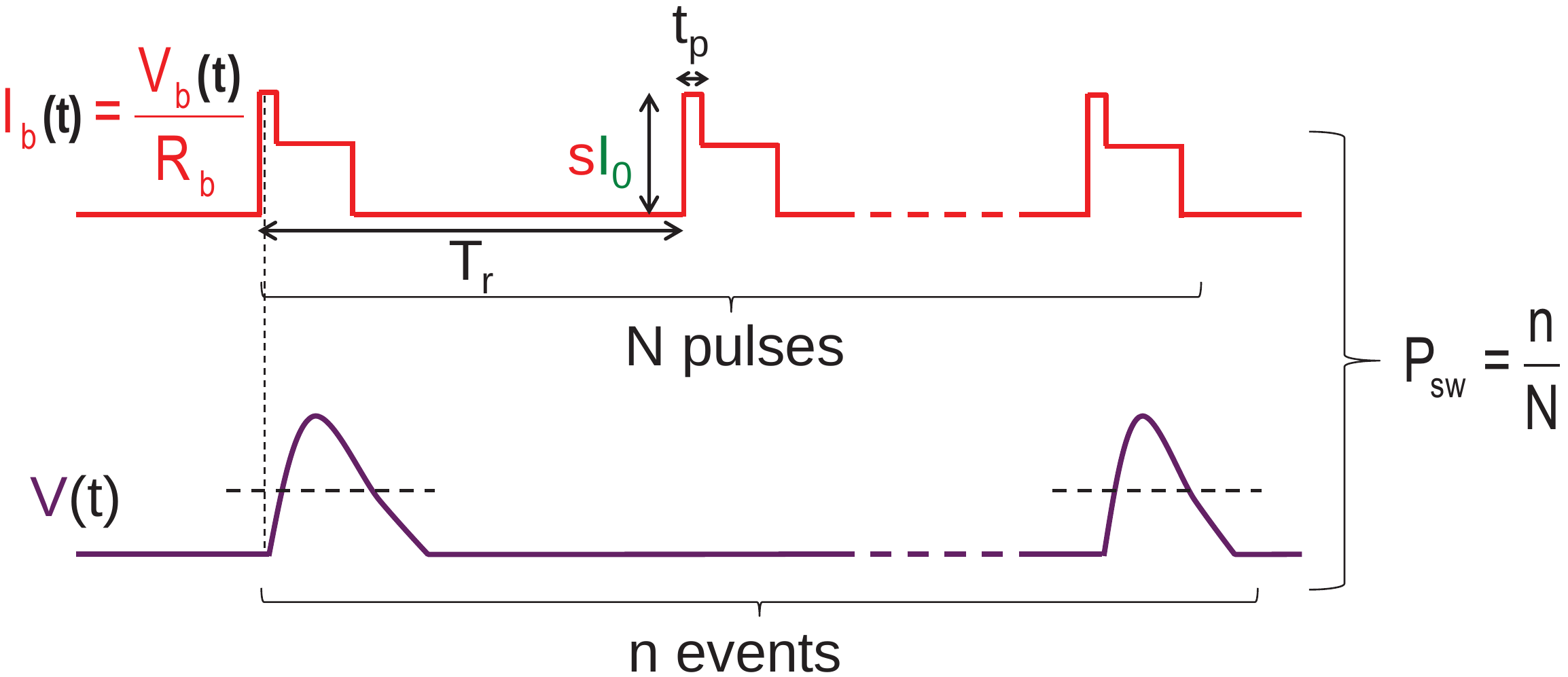} 
\par\end{centering}

\caption{Standard switching probability measurement scheme: a train of $N$
pulses with normalized height $sI_{0}$ and duration $t_{p}$ are
applied on the bias of the SQUID. The repetition rate is $1/T_{r}.$
The 40\% lower plateau following each pulse holds the voltage at a
finite value if switching has occurred, hence facilitating detection.
The number $n$ of voltage pulses resulting from switching events
is recorded by a counter detecting crossings through a threshold value
(dotted line). The switching probability is then $P_{sw}=\frac{n}{N}.$ }

\label{pulses} 
\end{figure}

Initially we performed switching measurement using a standard pulse
technique: a train of current pulses of normalized height $s=I/I_{0}$
is applied through the bias line, while monitoring the voltage across
the SQUID to reveal switching. The total number of voltage pulses
for a given current pulse train is recorded with a counter, see Fig.\,\ref{pulses}.
In order to have detectable voltage pulses even if switching occurs
at the end of the bias pulse, the actual bias pulse is followed by
a 40\% lower plateau that {}``holds'' the voltage at a finite value
for a time long enough for it to be detected. The height of this plateau
is such that the switching probability during its duration is negligible.
We used a bias pulse duration $t_{p}=$1\,\textmu{}s, and 5\,\textmu{}s-long
{}``hold'' plateaus. In general, the repetition period $T_{r}$
was 20\,\textmu{}s. The switching probability $P_{sw}\left(s\right)$
is obtained as the ratio of the number of measured voltage pulses
to the number of bias pulses (typically 10$^{4}$).%
\begin{figure}[h]
\begin{centering}
\includegraphics[clip,width=1\columnwidth]{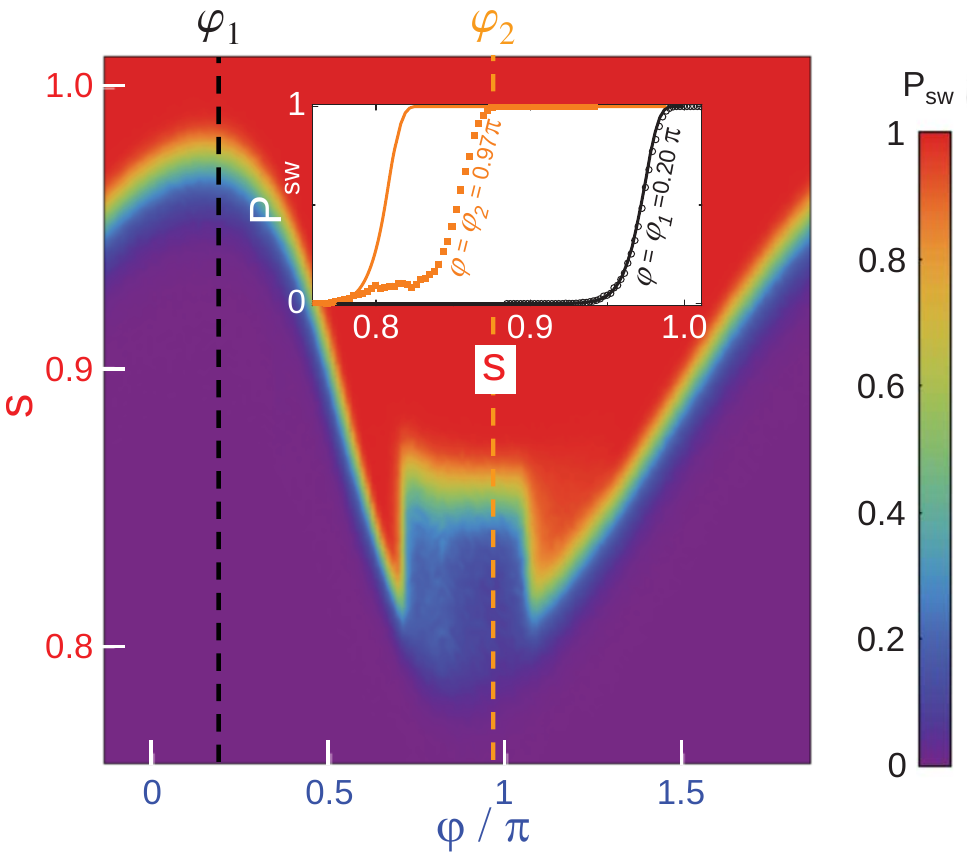} 
\par\end{centering}

\caption{Switching probability $P_{sw}$ measured using the standard method
(see Fig.~\ref{pulses}) as a function of flux phase $\varphi$ and
of the normalized pulse height $s$ for the SQUID with contact AC0
(\{0.95,0.45,0.10\}). Inset: $P_{sw}\left(s\right)$ for fluxes $\varphi_{1}=0.20\pi$
(black, open symbols) and $\varphi_{2}=0.97\pi$ (orange, full symbols).
Solid curves correspond to theoretical predictions.}

\label{Scurve}
\end{figure}
The inset of Fig.\,\ref{Scurve} shows the resulting switching probability
$P_{sw}$ as a function of the normalized pulse height $s$, for two
values of the flux phase corresponding to switching currents close
to its maximum and minimum values: $\varphi_{1}=0.20\pi$ (black symbols)
and $\varphi_{2}=0.97\pi$ (orange symbols). Whereas the shape of
the former (at $\varphi_{1}$) corresponds to what is usually found
on Josephson junctions, with a sharp variation of the probability
from 0 to 1 as the pulse height is increased, the latter (at $\varphi_{2}$)
presents an unusual long foot. The two types of behaviors are seen
in the main panel of Fig.\,\ref{Scurve}, a color plot of $P_{sw}$
with $\varphi$ and $s$ on the axes. Whereas the transition from
$P_{sw}\simeq0$ (purple) to $P_{sw}\simeq1$ (red) is abrupt for
most values of $\varphi,$ an anomalous region with an almost constant
intermediate step at $P_{sw}\simeq0.1$ (blue) is observed for $0.7\pi<\varphi<1.1\pi$.
In this region, the analysis of the histogram of time delays between
switching events shows that they are correlated, with bunches of switching
events and long {}``blind'' periods without any switching. This
behavior is reminiscent of the {}``blinking'' observed in the fluorescence
of molecules or quantum dots\cite{Dickson,Volkan}.

\begin{figure}[h]

\begin{centering}
\includegraphics[clip,width=0.7\columnwidth]{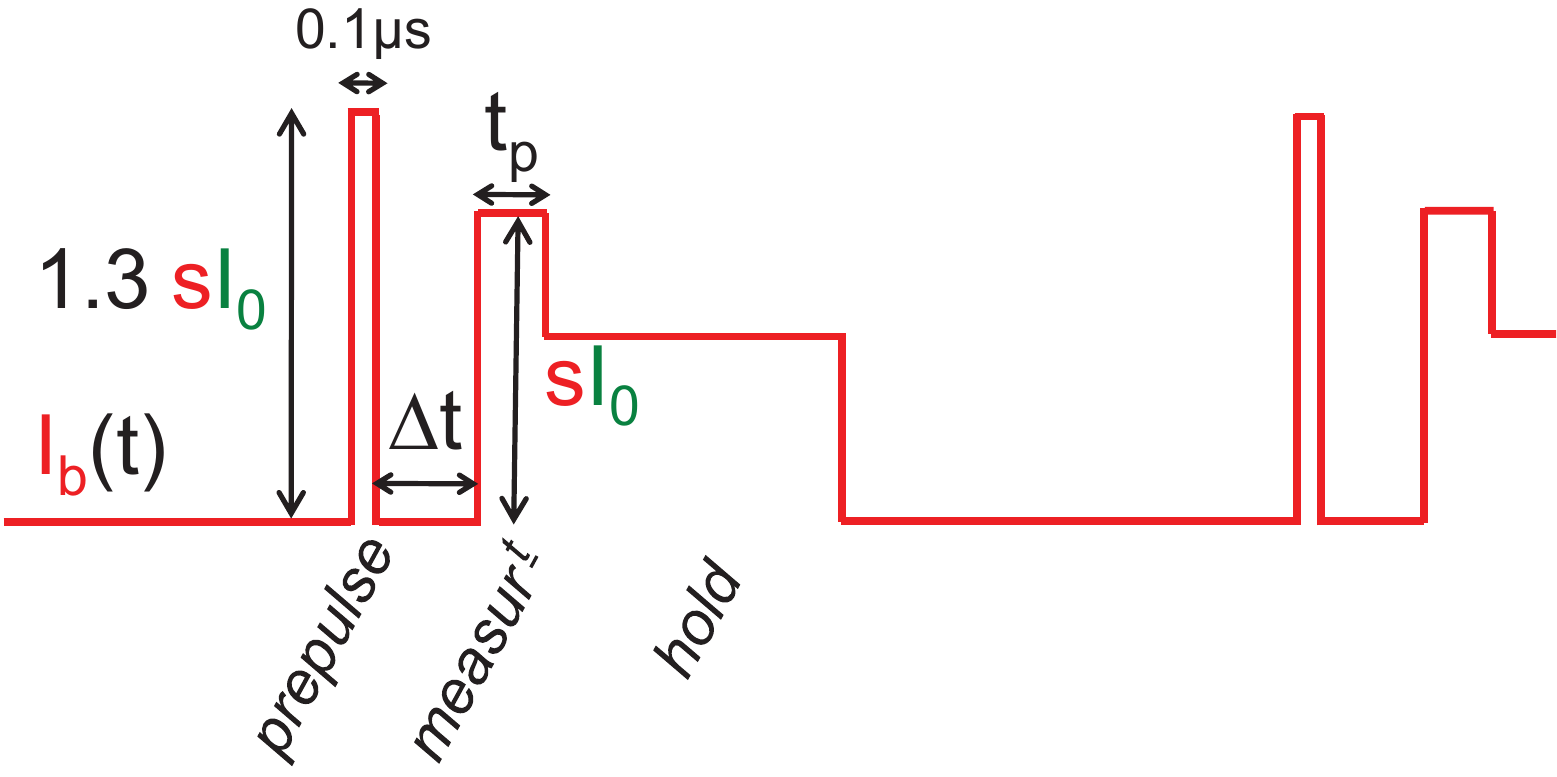} 
\par\end{centering}

\caption{Measurement scheme used in the following and for the data shown in
the Letter. Short prepulses 30\% higher than the actual measurement
pulses cause the system to always switch. With such pulses, successive
switching events are uncorrelated.}

\label{prepulse} 
\end{figure}

Correlations disappear when a strong bias pulse that forces the system
to switch (pulse height 30\% higher that the measurement pulse, see
Fig.\,\ref{prepulse}) is applied before each measurement pulse (of
course, the corresponding forced switching events are ignored in the
counting). During the prepulses, a voltage develops across the atomic
contact, and both the energy and the occupation of the Andreev levels
evolve at the Josephson frequency. When the current is reset to zero,
the Andreev levels are back to their energies $\pm E_{A}(\delta,\tau_{i})$,
but the system is left in an out-of-equilibrium situation, which on
average is always the same. The measurement pulse is applied a time
$\Delta t$ after this reset.\textbf{ }Using such {}``switching prepulses'',
the switching probability is independent of $T_{r}$, and the histograms
of time intervals between switching events do correspond to independent
processes.%
As reported in the main text, the corresponding curves $P_{sw}(s)$
then display a well defined intermediate plateau, as seen for flux
$\varphi_{2}$ in the inset of Fig.\,\ref{AC0}.

\section{Calculation of switching probability\label{sec:Calculation-of-switching.}}

We now give the details of the theoretical model used in the main
paper to describe the switching probability $P_{sw}(s,\varphi).$
It is related to the switching rate $\Gamma(s,\varphi)$ and to the
pulse duration $t_{p}$ by $P_{sw}=1-\exp(-\Gamma t_{p}).$ The phase
$\gamma$ across the Josephson junction is a dynamical variable governed
by a Langevin equation, equivalent to the one obeyed by the position
of a massive particle evolving in a {}``tilted washboard potential''\cite{Stewart,McCumber,Ambegaokar}.
The total potential of the SQUID is given by\cite{MLDR}:\begin{equation}
\begin{aligned}U(\gamma) & =-E_{J}\cos\gamma-E_{J}s\gamma+\sum_{i}c_{i}E_{A}(\gamma+\varphi,\tau_{i})\end{aligned}
\label{eq:potential}\end{equation}
where the first term is the Josephson energy of the tunnel junction,
with $E_{J}=I_{0}\varphi_{0}$, the second one is the energy arising
from the coupling to the bias source, and the last term is the total
Josephson coupling introduced by the atomic contact, which depends
on the configuration of the Andreev levels ($c_{i}=-1$ if channel
$i$ is in its ground state, $c_{i}=1$ in excited singlet, and $c_{i}=0$
in an odd configuration). Since $E_{J}/\Delta\simeq5.7\gg1,$ the
two first terms in Eq.~(\ref{eq:potential}) dominate, and the shape
of the tilted potential resembles that of a single Josephson junction,
with slightly modified barrier height $\Delta U(s,\varphi)$ and plasma
frequency $\omega_{p}(s,\varphi)$. As in Ref.~\onlinecite{MLDR},
switching can be fitted with thermal escape theory\cite{escape th},
with a rate $\Gamma=A\,\exp(-B)$ with $B=\Delta U/k_{B}T$ and $A\simeq\omega_{p}/2\pi.$
When the atomic contact is open, the plasma frequency in the tilted
potential is $\omega_{p}=\omega_{0}(1-s^{2})^{1/4}$, with $\omega_{0}\approx\sqrt{I_{0}/\varphi_{0}C_{JJ}}$,
$C_{JJ}=0.21\,$pF the Josephson junction capacitance estimated from
its area\cite{These QLM}, and $\Delta U\simeq(4\sqrt{2}/3)E_{J}(1-s)^{3/2}$
the barrier height\cite{escape MQT}. Fitting $P_{sw}(s)$ for the
Josephson junction alone gives $I_{0}=553.7\,\mathrm{nA}$ ($\omega_{0}/2\pi\approx15\,\mathrm{GHz),}$
and an effective temperature $T=100\,\mathrm{mK}$\cite{escapeT}.
When an atomic contact is formed, precise comparison between experiment
and theory is performed using a numerical determination of the barrier
height, and of the plasma frequency $\omega_{p}(s,\varphi)$ from
the semi-classical calculation of the energy levels in the actual
potential. More details on the SQUID potential are given in Appendix
A.%
\begin{figure}[h]
\begin{centering}
\includegraphics[clip,width=1\columnwidth]{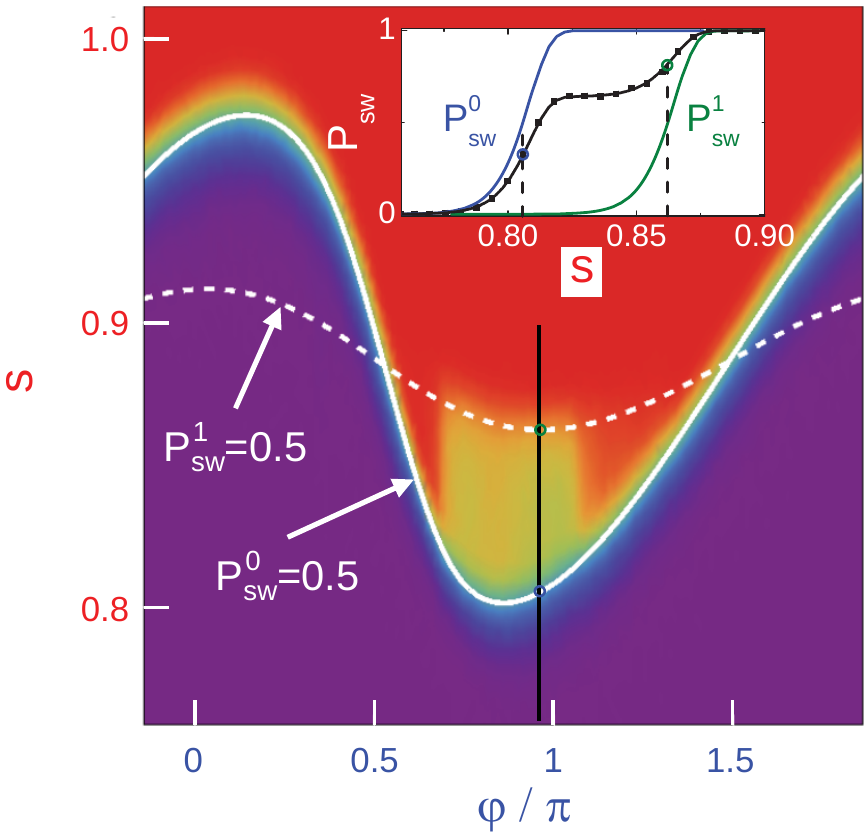} 
\par\end{centering}

\caption{Color plot of the switching probability $P_{sw}(s,\varphi)$ using
switching prepulses (see Fig.\,\ref{prepulse}, $\Delta t=0.5\,\text{\textmu s}$)
for the SQUID with contact AC0 with transmissions \{0.95,0.45,0.10\}.
The color scale is the same as for Fig.~\ref{Scurve}. The white
curves show the lines corresponding theoretically to $P_{sw}=0.5$
for the pristine contact (\{0.95,0.45,0.10\}, solid line) and for
the contact with the first channel poisoned (dashed line). Inset:
cut at $\varphi=0.97\pi$ (black line in the main panel). Black symbols:
$P_{sw}(s)$ as measured; solid curves: theoretical curves, $P_{sw}^{0}(s)$
(leftmost, blue online) for the pristine contact (\{0.95,0.45,0.10\})
and $P_{sw}^{1}(s)$ (rightmost, green online) for the poisoned contact,
i.e. without the contribution of the most transmitting channel; intermediate
line (black): fit of the data with the linear combination given by
Eq.~(\ref{eq:CL}) with $p=0.36$.}

\label{AC0}
\end{figure}

The predictions for the $P_{sw}(s)$ curves of the SQUID with contact
AC0 are shown as solid lines in the insets of Figs.\,\ref{Scurve}
and \ref{AC0}. Whereas the curve taken at $\varphi=\varphi_{1}=0.20\pi$
in Fig.\,\ref{Scurve} is well fitted by theory, the curve taken
at $\varphi=\varphi_{2}=0.97\pi$ has in common with theory only the
value of $s$ where the switching probability starts to rise. A central
point of the paper is that, when prepulses are used (Figs.\,\ref{AC0}),
the curve with a plateau can be very precisely accounted for by the
weighted sum of two $P_{sw}(s)$ curves corresponding to two different
configurations of the contact, as shown in the inset of Fig.\,\ref{AC0}:
$P_{sw}(s)=(1-p)P_{sw}^{0}(s)+pP_{sw}^{1}\left(s\right),$ with $p=0.36$,
as well as in the inset of Fig.~2 in the Letter. The first one, $P_{sw}^{0}(s),$
is the one predicted for the pristine contact; the second one, $P_{sw}^{1}(s),$
is that of the poisoned contact, \textit{i.e. }\textit{\emph{with
its most transmitting channel}} in an odd configuration. This is the
case in the whole flux region where the switching curves have an intermediate
plateau, as shown in the main panel of Fig.\,\ref{AC0}. As in the
Fig.~2 of the Letter, we compare the data with the two lines corresponding
to the equations $P_{sw}^{0,1}(s,\varphi)=0.5$. In the regions where
$P_{sw}(s)$ has a standard shape, $P_{sw}(s)$=0.5 occurs at the
position predicted for the pristine contact ($P_{sw}^{0}(s)$). In
the region $0.7\pi<\varphi<1.1\pi,$ $P_{sw}(s)$ has two steps (which
appear in the figure as color gradients): one occurs at the position
where $P_{sw}^{0}(s,\varphi)=0.5$ (solid line), the other one at
the position where $P_{sw}^{1}(s,\varphi)=0.5$ (dashed line), showing
that \begin{equation}
P_{sw}(s,\varphi)=(1-p\left(\varphi\right))P_{sw}^{0}(s,\varphi)+p\left(\varphi\right)P_{sw}^{1}\left(s,\varphi\right).\label{eq:CL}\end{equation}

\begin{figure*}
\begin{centering}
\includegraphics[clip,width=2\columnwidth]{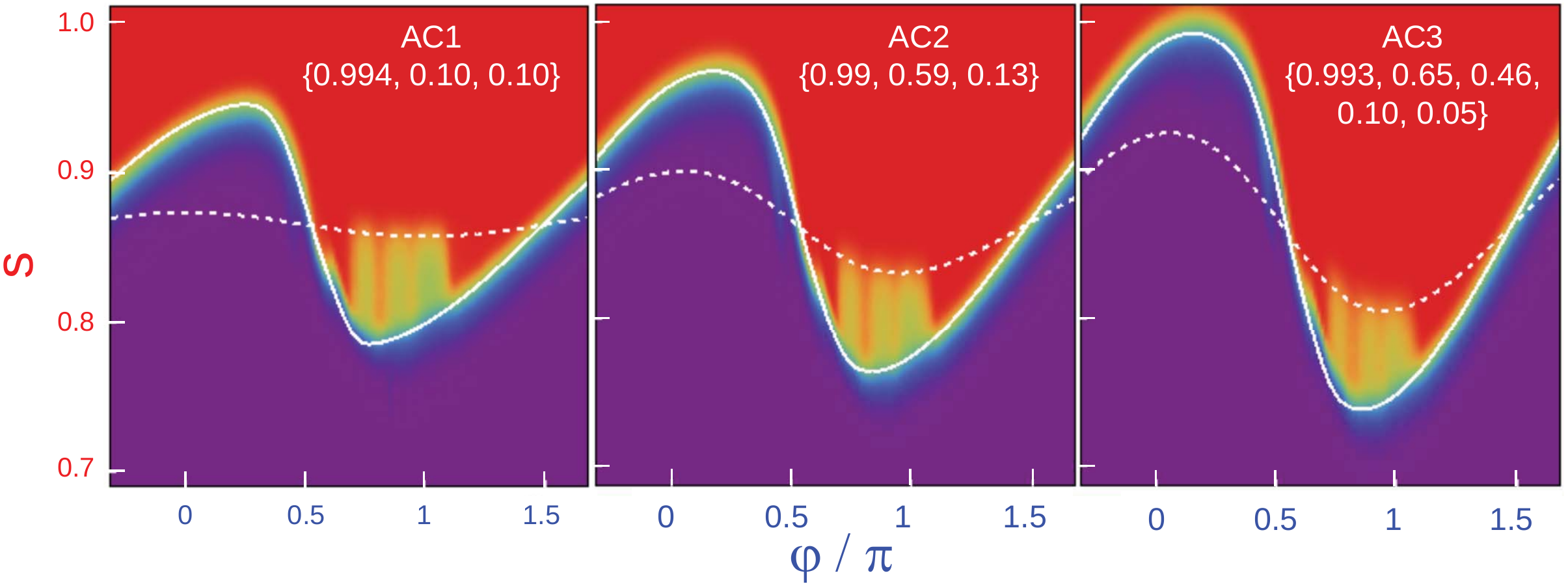} 
\par\end{centering}

\caption{Color plot of the switching probability $P_{sw}(s,\varphi)$ using
switching prepulses (see Fig.\,\ref{prepulse}, with $\Delta t=2\,\mathrm{\mu s}$)
for SQUIDs made with three different atomic contacts, each having
one channel with a transmission close to 1: AC1, \{0.994,0.10,0.10\}
(same data as in the Letter); AC2: \{0.998,0.59,0.13\} and AC3: \{0.993,0.65,0.46,0.10,0.05\}.
The white curves correspond theoretically to $P_{sw}=0.5$ for the
pristine contact (solid line) and for the poisoned contact (dashed
line).}

\label{3Dplots} 
\end{figure*}

In Fig.~\ref{3Dplots}, similar comparisons are presented for three
other SQUIDs formed with atomic contacts having one channel almost
perfectly transmitting and the other ones with transmissions lower
than 0.7. There again, in a broad phase range around $\pi,$ $P_{sw}(s)$
shows a plateau delimited by the predictions for the pristine (solid
line) and for the poisoned configurations, i.e. with the more transmitting
contact in an odd configuration (dashed line).

\section{Measurements of poisoning dynamics}

\subsection{Method}

\begin{figure}[h]
\begin{centering}
\includegraphics[clip,width=1\columnwidth]{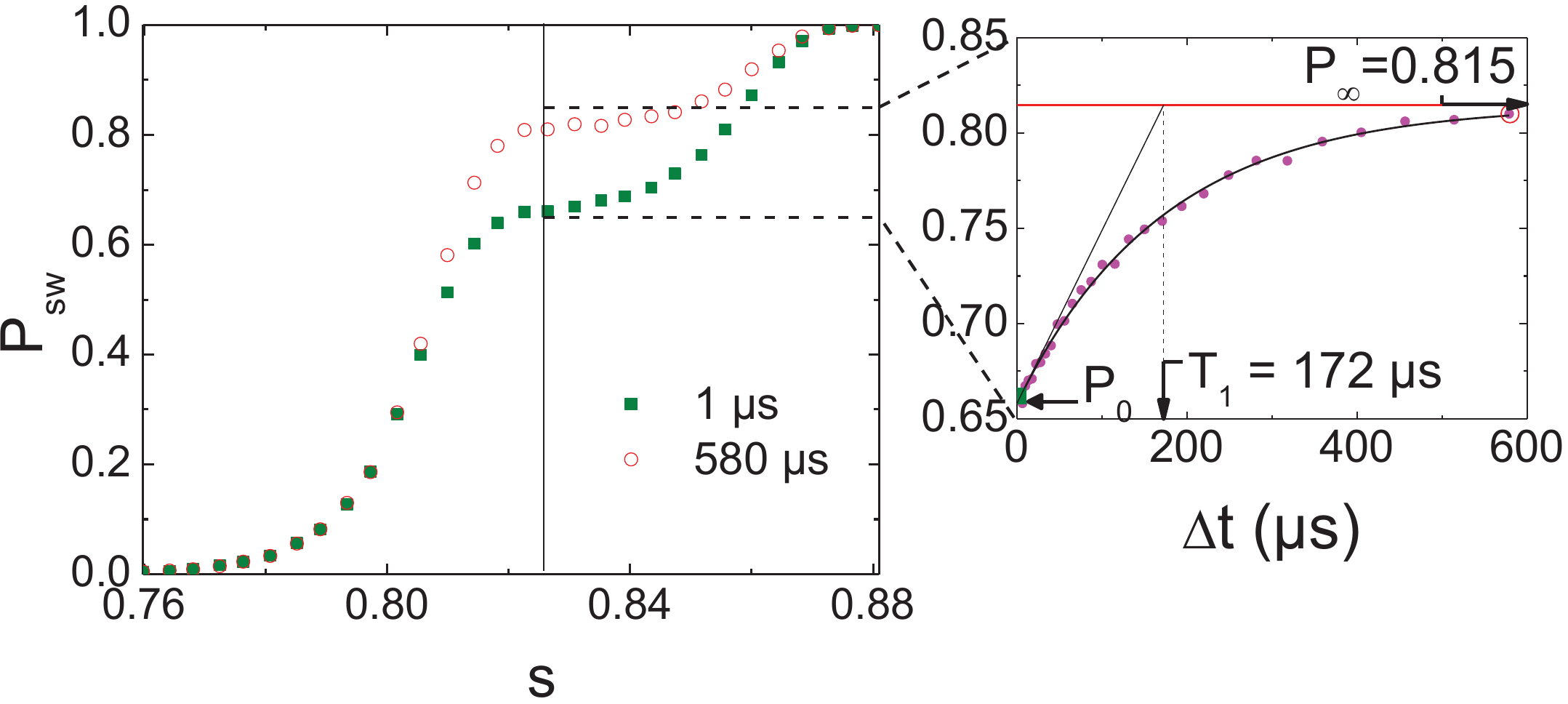} 
\par\end{centering}

\caption{Left panel: Switching probability of contact \{0.91, 0.62, 0.15\}
as a function of bias pulse height at fixed $\varphi$, for short
($\Delta t=1\,\text{\textmu}s$, green solid squares) and long ($\Delta t=580\,\text{\textmu}s,$
red open circles) delay between the prepulse and the measurement pulse.
Right panel: Dots: Evolution with $\Delta t$ of the plateau height,
measured at $s=0.826$ (vertical line on the left panel). Solid line:
exponential fit with $P_{sw}(\Delta t)=P_{\infty}+\left(P_{0}-P_{\infty}\right)\exp\left(-\Delta t/T_{1}\right).$}

\label{relax} 
\end{figure}

The previous experiments demonstrate that, in a certain parameter
region, the system has a finite probability $p$ to trap a quasiparticle
and end in an odd configuration after the prepulse. In the following,
we describe the experiments exploring the dynamics of trapping and
untrapping, by varying the delay $\Delta t$ between the prepulse
and the measurement pulse. Data illustrative for the method, taken
on a contact with transmissions \{0.91, 0.62, 0.15\}, are shown Fig.\,\ref{relax}:
in a flux region exhibiting poisoning, the left panel shows $P_{sw}(s)$
for a short and a long delay $\Delta t$, and the right one the complete
dependence of $P_{sw}$ on $\Delta t$ at a bias value $s=0.826$,
for which $P_{sw}^{0}\approx1$ and $P_{sw}^{1}\approx0.$ The data
are well fitted with an exponential dependence $P_{sw}(\Delta t)=P_{\infty}+(P_{0}-P_{\infty})\exp(-\Delta t/T_{1}).$
One can then extract the initial poisoning just after the prepulse
$p_{0}=1-P_{0}$, the asymptotic value at long times $p_{\infty}=1-P{}_{\infty}$,
and the relaxation time $T_{1}$.

As discussed in the Letter, the dynamics of quasiparticle poisoning
is strongly phase dependent. Measuring this phase dependence is however
not trivial, because during the pulse sequence shown in Fig.~\ref{prepulse},
$\delta$ is not constant: it takes a value close to $\varphi$ during
$\Delta t,$ then reaches $\sim\varphi+\gamma\left(s_{sw}^{JJ}\right)$
during the measurement pulse ($\gamma\left(s_{sw}^{JJ}\right)\simeq0.3\pi$).
As shown below, relaxation is very fast when $\delta>1.3\pi,$ so
that data taken at $\varphi>\pi$ are dominated by the relaxation
at measurement.

\begin{figure}[h]
\begin{centering}
\includegraphics[clip,width=0.9\columnwidth]{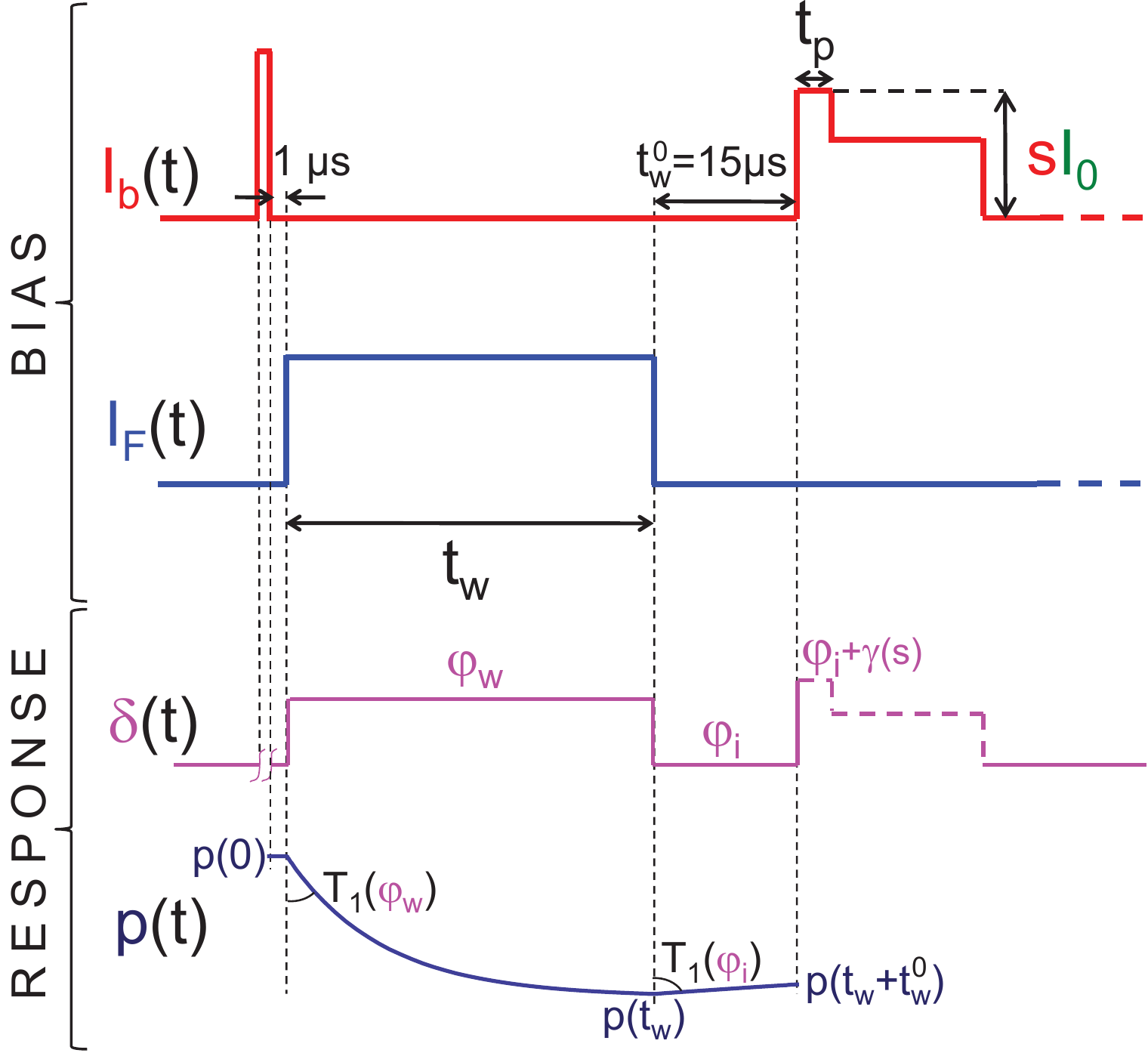} 
\par\end{centering}

\caption{Pulse sequence used to measure the phase dependence of the relaxation
process. The first line corresponds to the signal applied to the current
bias, the second one to the fast flux line. The two last lines sketch
the corresponding evolution of the phase $\delta$ across the atomic
contact and of the poisoning probability $p\left(t\right)$. The prepulse
causes systematic switching, and the {}``running away'' of the phase
is not represented. The dotted line in $\delta(t)$ during the measurement
pulse indicates that either the SQUID switches and the phase runs
away, or it does not switch and $\delta$ simply follows the bias
current. The prepulse and the measurement pulse are always applied
at the same flux phase $\varphi_{i}.$ The probability $p\left(t\right)$
starts from $p\left(0\right)$ after the prepulse, a value that depends
on the flux $\varphi_{i}$; a flux phase $\varphi_{w}$ is then applied
during a time $t_{w}$, and $p(t)$ relaxes exponentially with a time
constant $T{}_{1}\left(\varphi_{w}\right)$ towards $p_{\infty}\left(\varphi_{w}\right)$,
reaching $p\left(t_{w}\right).$ In the last step, the phase flux
is restored to $\varphi_{i}$ during $t_{w}^{0}=15\,\text{\textmu s}$
and $p\left(t\right)$ evolves with the time constant $T{}_{1}\left(\varphi_{i}\right)$
towards $p_{\infty}\left(\varphi_{i}\right)$, reaching finally $p(t_{w}+t_{w}^{0}),$
the actual value accessed by the measurement (the schematic corresponds
to a situation where $p\left(t_{w}\right)<p_{\infty}\left(\varphi_{i}\right)$,
hence $p\left(t\right)$ increases in the last step). In the time
interval between the prepulse and the measurement pulse, the bias
current $I_{b}$ is zero, and the phase $\delta$ across the atomic
contact is equal to the flux phase: $\delta=\varphi_{w}$ during $t_{w}$,
then $\delta=\varphi_{i}$ during $t_{w}^{0}.$}

\label{Fluxpulse} 
\end{figure}

To correctly measure the phase dependence of the relaxation, we have
therefore elaborated a different protocol, shown in Fig.~\ref{Fluxpulse}.
We set the flux imposed by the external coil to a value $\varphi_{i}$
such that $p_{0}$ is far from 0, and that the relaxation times at
$\varphi_{i}$ and at $\varphi_{i}+\gamma\left(s_{sw}^{JJ}\right)$
are long, so that relaxation during the measurement can be neglected.
We first take a switching curve $P_{sw}\left(s\right)$ and fit it
with a weighted sum of two shifted curves as in the inset of Fig.~\ref{AC0}.
We then fix a working point $s_{0}$ corresponding to the intermediate
plateau, and determine the switching probabilities corresponding to
the pristine contact $P_{sw}^{0}\left(s_{0}\right)$ and to the poisoned
contact $P_{sw}^{1}\left(s_{0}\right).$ For all the data presented
in the following, the working point was chosen such that $P_{sw}^{0}\left(s_{0}\right)=1.00$
and $0.01\leq P_{sw}^{1}\left(s_{0}\right)\leq0.12$ (except for the
data taken at higher temperatures, where the rounding of the curves
becomes comparable to their width, and $P_{sw}^{1}\left(s_{0}\right)$
can be as high as 0.32). The probability $p$ to be in an odd configuration
is then inferred from \begin{equation}
P_{sw}(s_{0})=(1-p)P_{sw}^{0}(s_{0})+pP_{sw}^{1}\left(s_{0}\right),\label{eq:Ptop}\end{equation}
which is slighlty more precise than the identification of $p$ with
$1-P_{sw}(s_{0})$. A measurement similar to that presented in Fig.~\ref{relax}
is then used to characterize the system at flux $\varphi_{i}:$\begin{equation}
p(\varphi_{i},\Delta t)=\mathcal{E}_{T_{1}\left(\varphi_{i}\right)}^{p_{\infty}\left(\varphi_{i}\right)}\left(p_{0}\left(\varphi_{i}\right),\Delta t\right)\end{equation}
where we have introduced the function $\mathcal{E}_{T_{1}}^{p_{\infty}}\left(p_{0},t\right)$
accounting for an exponential variation starting from $p_{0}$ and
during time $t,$ with characteristic time $T_{1}$ and asymptotic
value $p_{\infty}:$ \begin{equation}
\mathcal{E}_{T_{1}}^{p_{\infty}}\left(p_{0},t\right)\equiv p_{\infty}+(p_{0}-p_{\infty})\exp\left(-t/T_{1}\right).\end{equation}
Using a dc flux pulse applied through the fast flux line between the
prepulse and the measurement pulse, the flux phase is changed to a
value $\varphi_{w}$ for a time $t_{w}$ (in practice, we leave a
1~\textmu{}s delay between the prepulse and the flux pulse to let
the system stabilize; we also leave a delay of $t_{w}^{0}=$15~\textmu{}s
between the flux pulse and the measurement pulse to get rid of ringing
effects after the fast flux pulse). Since the bias current is 0 during
this dc flux pulse, the phase across the atomic contact is then simply
$\delta=\varphi_{w}$ \cite{phaseJJ}.

From the switching probability $P_{sw}\left(t_{w}+t_{w}^{0}\right)$
we calculate using Eq.~(\ref{eq:Ptop}) the probability $p\left(t_{w}+t_{w}^{0}\right)$
to be in an odd configuration after the complete sequence. We observe
that $p\left(t_{w}+t_{w}^{0}\right)$ is not an exponential function
of $t_{w}$. The reason is that this probability results from two
exponential relaxations with different parameters, as illustrated
at the bottom of Fig.~\ref{Fluxpulse}. The initial value $p(0)=p_{0}\left(\varphi_{i}\right)$
results from the prepulse applied at phase flux $\varphi_{i}$ (the
function $p_{0}\left(\varphi_{i}\right)$ is discussed farther). Follows
an exponential evolution at phase flux $\varphi_{w}$ during $t_{w}$,
leading to \begin{equation}
p(t_{w})=\mathcal{E}_{T_{1}\left(\varphi_{w}\right)}^{p_{\infty}\left(\varphi_{w}\right)}\left(p\left(0\right),t_{w}\right).\label{eq:pinterm}\end{equation}
During the last $t_{w}^{0}=15\,\text{\textmu s}$, the phase flux
is $\varphi_{i}$, so that \begin{equation}
p\left(t_{w}+t_{w}^{0}\right)=\mathcal{E}_{T_{1}\left(\varphi_{i}\right)}^{p_{\infty}\left(\varphi_{i}\right)}\left(p\left(t_{w}\right),t_{w}^{0}\right).\label{eq:pfinal}\end{equation}
Since the parameters $p(0),$ $T_{1}\left(\varphi_{i}\right)$ and
$p_{\infty}\left(\varphi_{i}\right)$ have been determined in the
first measurement without the flux pulse, Eq.~(\ref{eq:pfinal})
can be used to deduce $p\left(t_{w}\right)$ from $p\left(t_{w}+t_{w}^{0}\right).$
The function $p\left(t_{w}\right)$ is then an exponential, as expected,
and its fit with Eq.~(\ref{eq:pinterm}) yields the asymptotic poisoning
$p_{\infty}\left(\delta\right)$ and the relaxation time $T_{1}\left(\delta\right).$ 

Using this more complex procedure that involves the fast flux line
to apply dc flux pulses, the relaxation can be probed with the prepulse
and the measuring pulse both acting always at the same flux. The initial
value of the poisoning probability is therefore always the same, and
the working point can be chosen such that relaxation during the measurement
pulse plays no role (note that, as long as this last criterium is
observed, $p_{\infty}\left(\delta\right)$ and $T_{1}\left(\delta\right)$
do not depend on the working point).

\subsection{Temperature dependence}

\begin{figure}[h]
\begin{centering}
\includegraphics[clip,width=1\columnwidth]{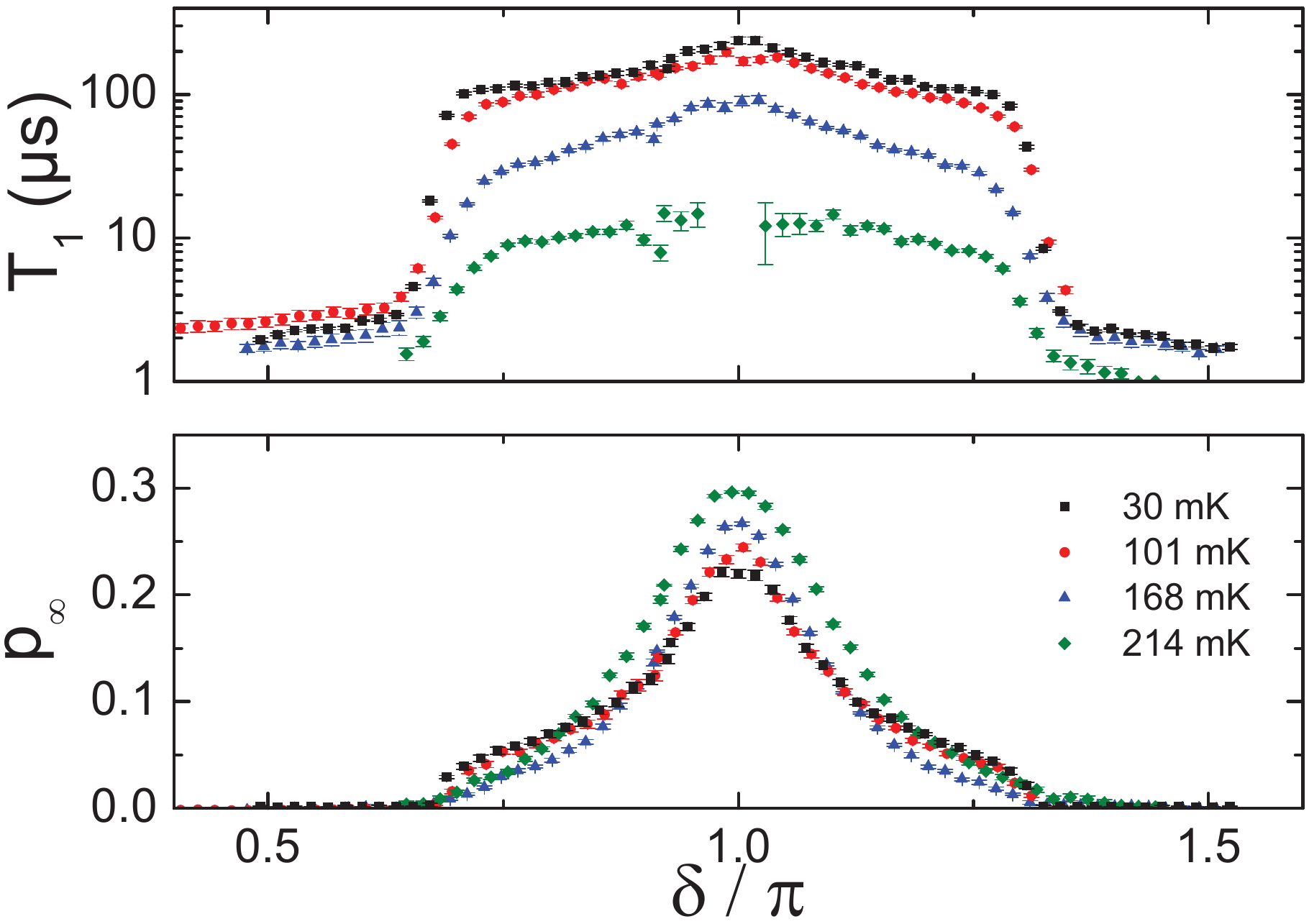}
\par\end{centering}

\caption{Relaxation time $T_{1}$ and asymptotic poisoning probability $p_{\infty},$
as a function of the phase $\delta=\varphi_{w}$ applied during the
dc flux pulse imposed for a time $t_{w}$ between the prepulse and
the measurement pulse. The data are taken on the atomic contact AC1
(\{0.994,0.10,0.10\}), and at four different temperatures $T$ indicated
in the figures.}

\label{pollution1} 
\end{figure}

The functions $p_{\infty}\left(\delta\right)$ and $T_{1}\left(\delta\right)$
at various temperatures, are shown in Fig.\,\ref{pollution1} for
contact AC1. Whereas $p_{\infty}\left(\delta\right)$ hardly changes,
$T_{1}\left(\delta\right)$ drops rapidely with temperature, and relaxation
is almost instantaneous above 220~mK. %

\subsection{Revisiting $P_{sw}\left(s,\varphi\right)$ data}

Coming back to the results presented in the left panel of Fig.~\ref{3Dplots},
the boundaries of the phase range in which poisoning occurs can be
understood in the light of Fig.~\ref{pollution1}: the left boundary
corresponds to the phase $\varphi_{L}$ at which $T_{1}\left(\varphi_{L}\right)$
becomes comparable with $\Delta t=2\,\mathrm{\mu s}$, leading to
significant relaxation in the time interval between the prepulse and
the measurement pulse. The nature of the right boundary is different:
it corresponds to the phase $\varphi_{R}$ at which $T_{1}\left(\gamma\left(s\right)+\varphi_{R}\right)$
becomes comparable with $t_{p}=1\,\mathrm{\mu s}$ (we recall that
$\gamma\left(s\right)$ is the phase across the Josephson junction
during the measurement pulse), leading to significant relaxation during
the measurement pulse. In contrast with the left boundary, the position
of the right one depends slightly on $s$ through $\gamma\left(s\right),$
which explains that it is slightly tilted. Hence, in the simplest
procedure where the flux is the same during the whole sequence, the
effect of relaxation during the measurement pulse becomes predominent
for $\varphi>\pi,$ explaining why the intervals on which poinsoning
is observed in Fig.~\ref{3Dplots} is not centered at $\pi.$

\subsection{Revisiting data without prepulses}

The anomalous statistics of the time intervals between switching events
mentioned above finds also an explanation. The data presented in Fig.~\ref{Scurve}
are taken without prepulse, so that if poisoning occurs after a measurement
pulse, the system remains poisoned during a time $T_{1}$ in average,
and switching is suppressed, giving rise to long {}``blind'' periods.
When the system unpoisons, switching occurs again, and, the probability
to get poisoned again being rather small, several switching events
occur in a row. When the repetition period $T_{r}$ is increased,
the probability to escape from a poisoned configuration before the
next measurement increases. And since the switching rate is larger
when the system is not poisoned, the average switching probability
increases, as observed.

\subsection{Raw data on several contacts}

\begin{figure}
\begin{centering}
\includegraphics[clip,width=1\columnwidth]{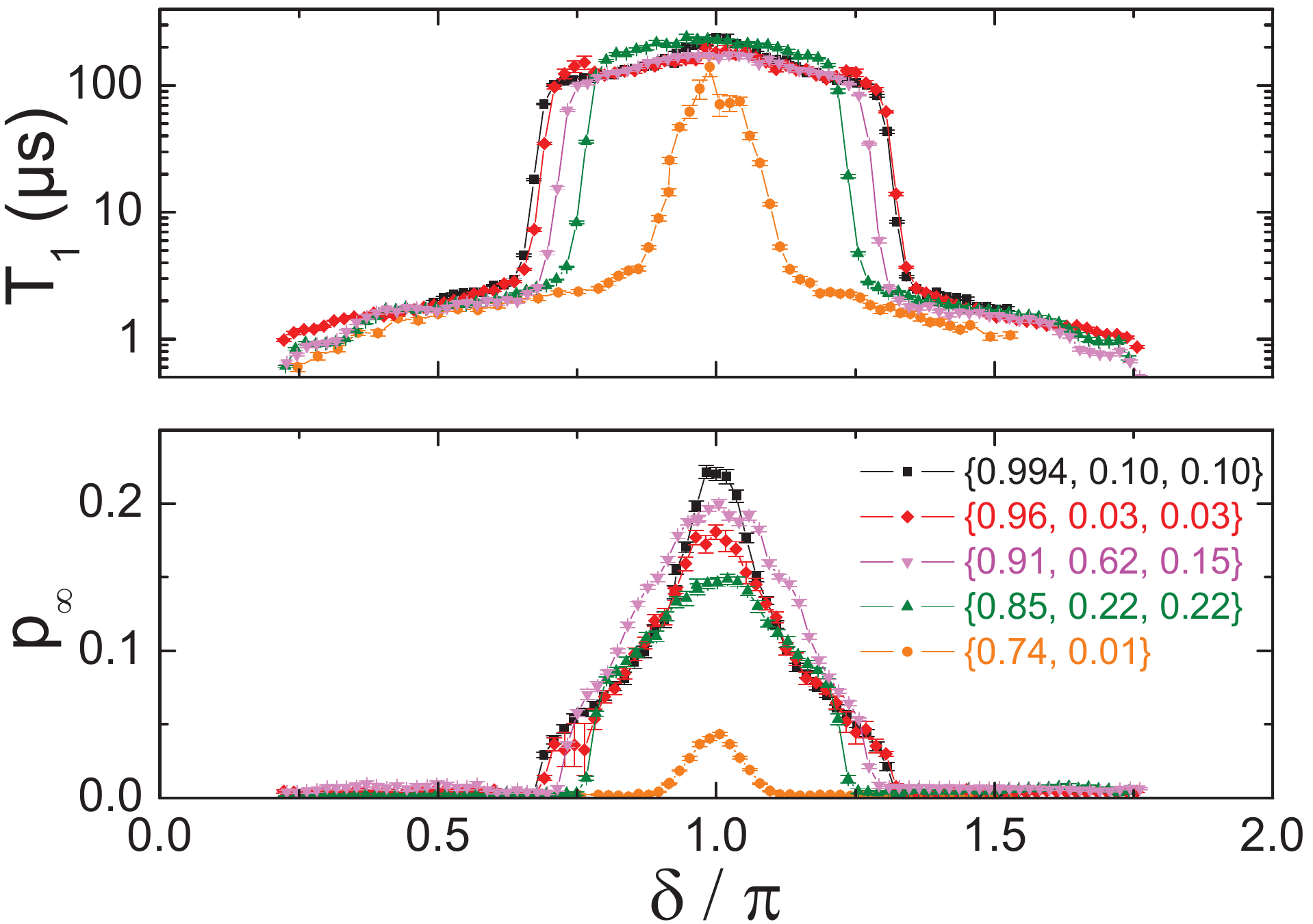} 
\par\end{centering}

\caption{Relaxation time $T_{1}$ and asymptotic poisoning probability $p_{\infty},$
measured at $T=20$\,mK, for five different atomic contacts with
the shown transmissions.}

\label{pollution2} 
\end{figure}

Similar data taken on a variety of atomic contacts are shown in Fig.\,\ref{pollution2}.
The phase interval in which poisoning occurs reduces when the transmission
of the most transmitting channel diminishes. The same data are replotted
as a function of $E_{A}$ in Fig.~4 of the Letter. %

\subsection{Multiple poisoning}

\begin{figure}
\begin{centering}
\includegraphics[clip,width=0.8\columnwidth]{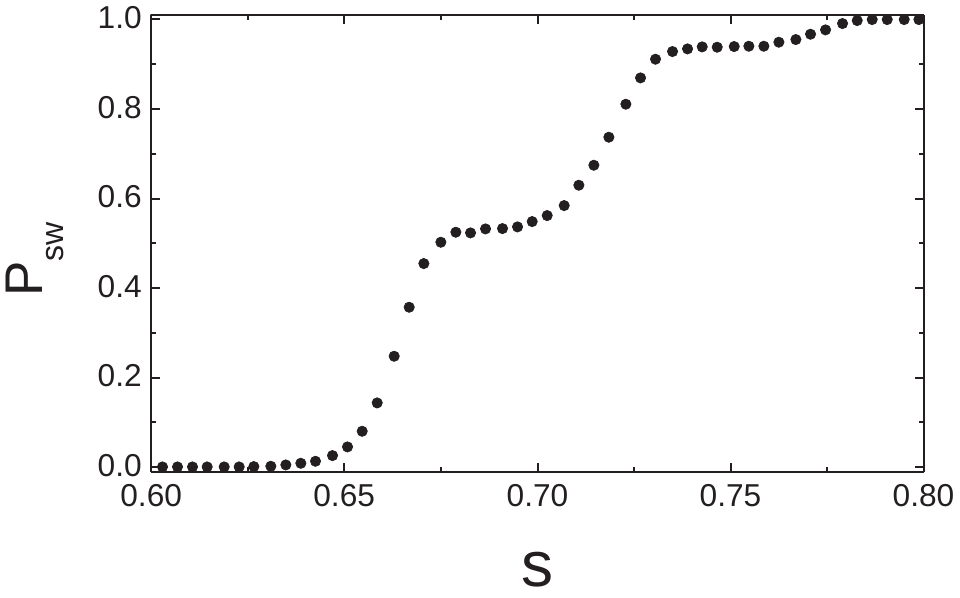} 
\par\end{centering}

\caption{Measured switching probability as a function of $s$ for a contact
with two well transmitting channels: \{0.96, 0.95, 0.60, 0.34, 0.30,
0.29, 0.27, 0.26, 0.24, 0.2\}, taken at a flux within the poisoned
region. The first plateau ($P_{sw}\simeq0.55$) can be attributed
to situations where one of the two channels with transmission $\sim0.95$
is poisoned, while the second one corresponds to situations where
both are poisoned.}

\label{double} 
\end{figure}

In a contact with more than one well transmitting channel, poisoning
can affect several channels at once, as shown in Fig.~\ref{double}
where the switching probability presents two intermediate plateaus.
The first one near 0.5 corresponding to either one of the two first
channels being poisoned (they have very similar transmissions), the
second one at 0.95 to the situation in which both are.

\subsection{Initial poisoning}

\begin{figure}
\begin{centering}
\includegraphics[clip,width=0.9\columnwidth]{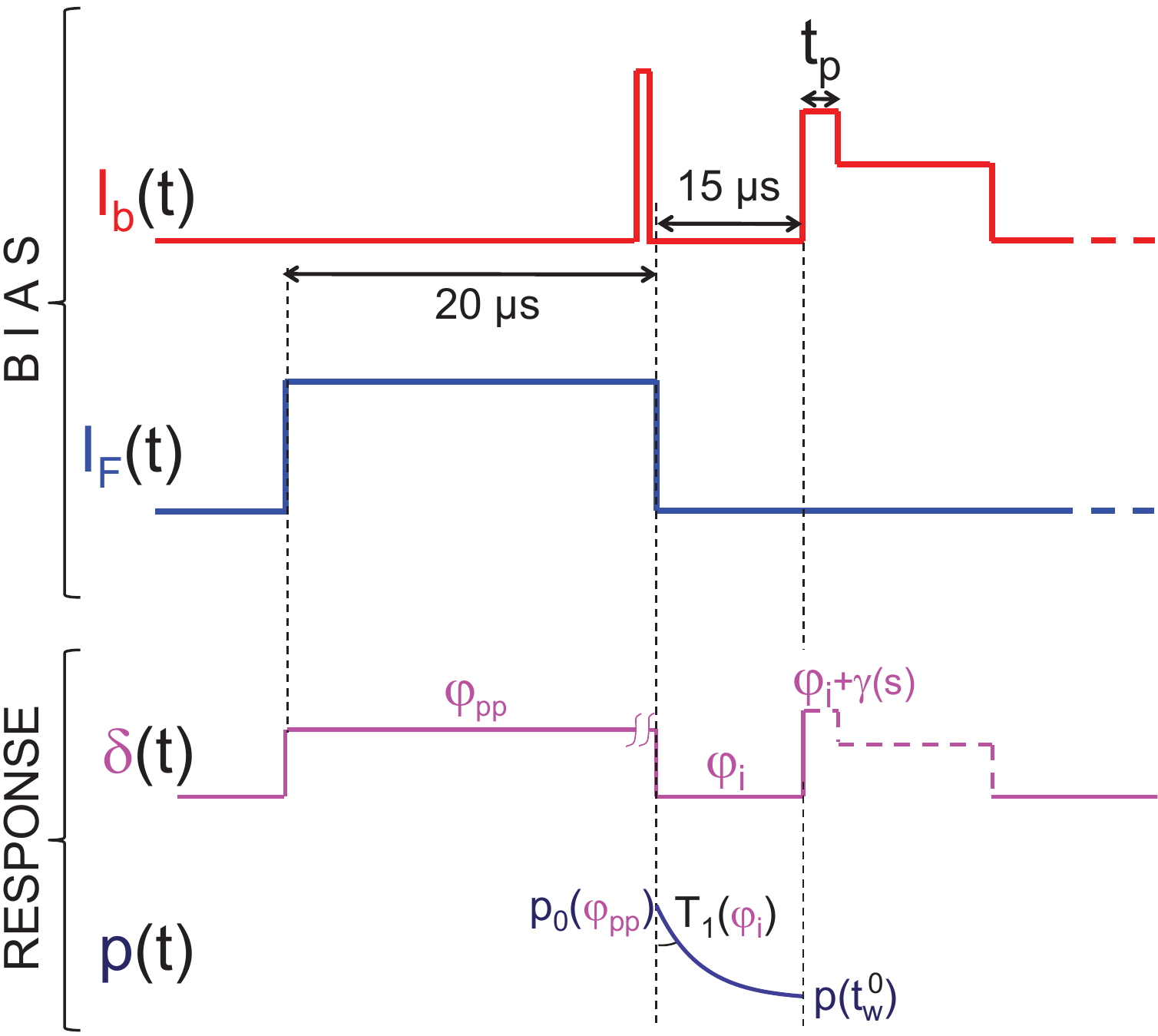} 
\par\end{centering}

\caption{Pulse sequence used to measure the phase dependence of the initial
poisoning probability $p_{0}$. The first line corresponds to the
bias current, and the second one to the fast flux line. The third
line sketches the evolution of the phase $\delta$ across the atomic
contact, and the last one that of the poisoning probability. The quantity
that is studied is the poisoning probability $p_{0}\left(\varphi_{pp}\right)$
just after the prepulse, which depends on the phase flux $\varphi_{pp}$
applied to the contact during the prepulse. Note that the measuring
pulse gives access to $p\left(t_{w}^{0}\right),$ which includes relaxation
at flux $\varphi_{i}$ during the last $t_{w}^{0}=15\,\text{\textmu}s.$
This relaxation is taken into account to obtain $p_{0}\left(\varphi_{pp}\right)$.}

\label{Fluxpulse2} 
\end{figure}

\begin{figure}
\begin{centering}
\includegraphics[clip,width=0.9\columnwidth]{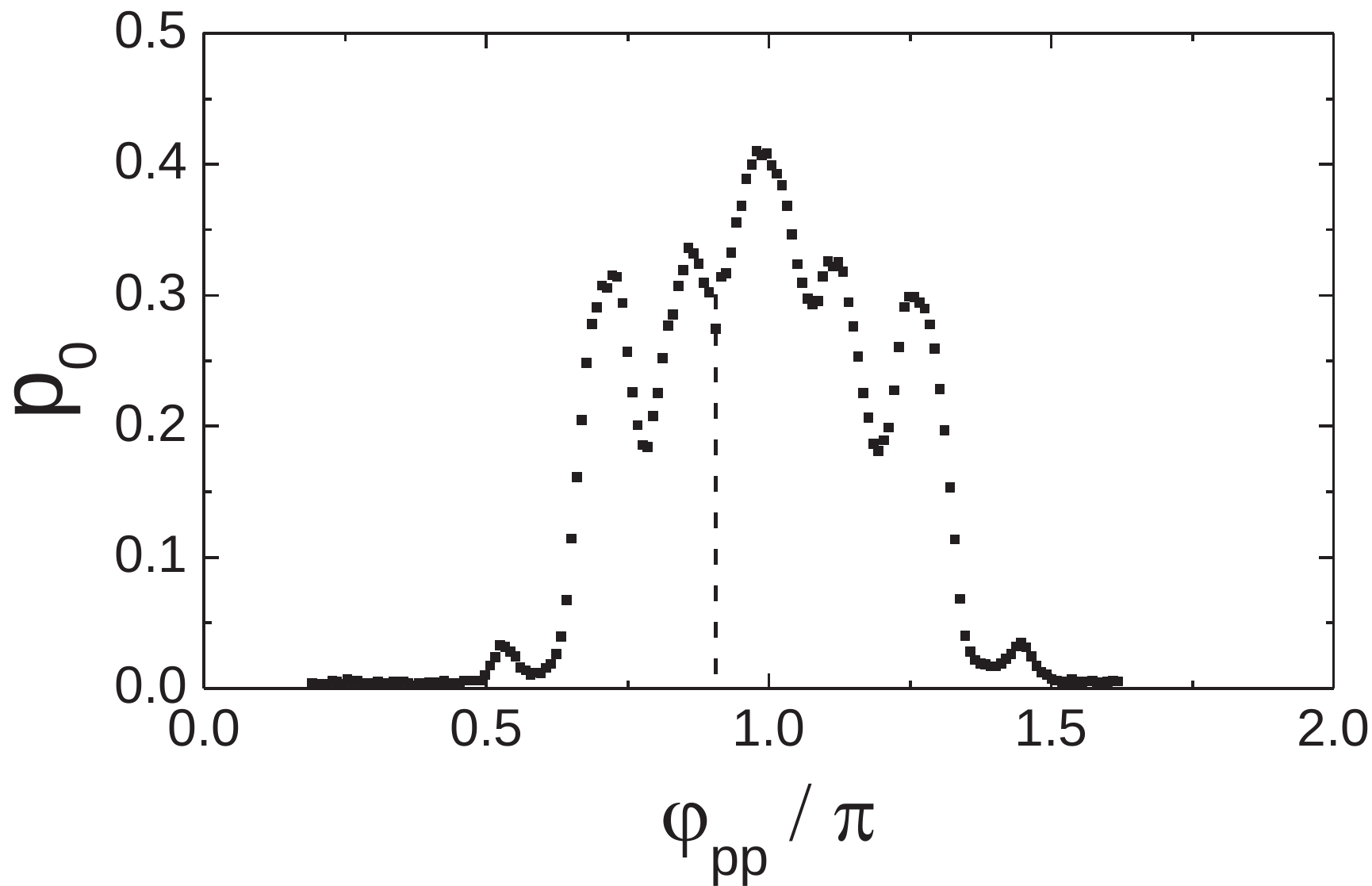} 
\par\end{centering}

\caption{Initial poisoning $p_{0}$ as a function of the phase $\delta=\varphi_{pp}$
across the atomic contact AC2 at the end of the prepulse. The pulse
sequence used to gather this data-set is shown in Fig.~\ref{Fluxpulse2}.
The dashed line indicates the value of the phase $\delta$ across
the atomic contact when only the dc flux $\varphi_{i}$ is applied.}

\label{p0} 
\end{figure}

We have measured the initial poisoning $p_{0}$ as a function of the
phase $\delta$ across the atomic contact AC1 (\{0.994,0.10,0.10\}).
The measurement protocol is shown in Fig.\,\ref{Fluxpulse2}: the
flux phase $\varphi_{pp}$ applied till the end of the prepulse is
varied, then the flux is reset to $\varphi_{i}$. Here again, the
effect of relaxation at $\varphi_{i}$ during the 15~\textmu{}s between
the end of the flux pulse and the measurement pulse has been corrected
for. The result of the measurement is displayed in Fig.~\ref{p0}
(for the chosen value of $\varphi_{i},$ indicated with a dashed line,
the relaxation before measurement is characterized by $T_{1}=167\,\text{\textmu s}$
and $p_{\infty}=0.12$). This irregular pattern is responsible for
the vertical stripes in the data shown in Fig.~\ref{3Dplots}. Similar
patterns, with slight dependence on the duration of the prepulse and
on the rate of decay of the current at the end of the prepulse, were
found on the other contacts, but the influence of the different parameters
could not be deconvolved from the data. 


\section{Quasiparticle poisoning antidote\label{sec:Quasiparticle-poisoning-antidote}}

\begin{figure}
\begin{centering}
\includegraphics[clip,width=0.8\columnwidth]{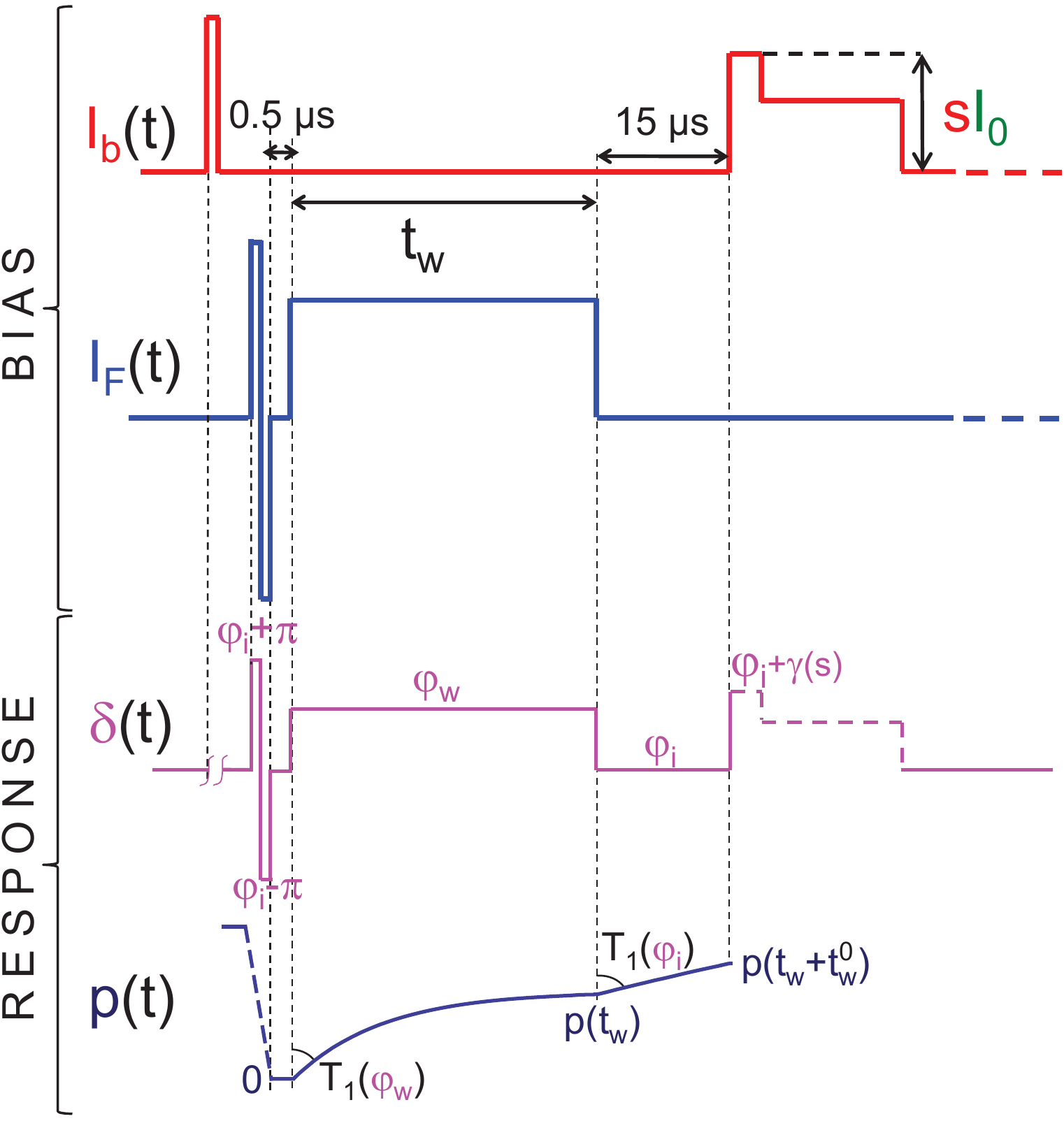} 
\par\end{centering}

\caption{Pulse sequence used to measure the phase dependence of the relaxation
process after the application of {}``antidote pulses''. The first
line corresponds to the current bias, and the second one to the fast
flux line. The third line sketches the evolution of the phase $\delta$
across the atomic contact, and the last one that of the poisoning
probability $p\left(t\right)$. The antidote pulses are pairs of dc
flux pulses applied in both directions, with an amplitude corresponding
to a phase excursion of $\pi$ on the phase $\delta.$ Their total
duration was 0.5~\textmu{}s; they are applied 1.5~\textmu{}s after
the prepulse. During the pair of antidote pulses, the phase visits
regions where a trapped quasiparticle escapes rapidly, and the poisoning
probability decays to 0.}

\label{clean pulse} 
\end{figure}

\begin{figure}
\begin{centering}
\includegraphics[clip,width=0.8\columnwidth]{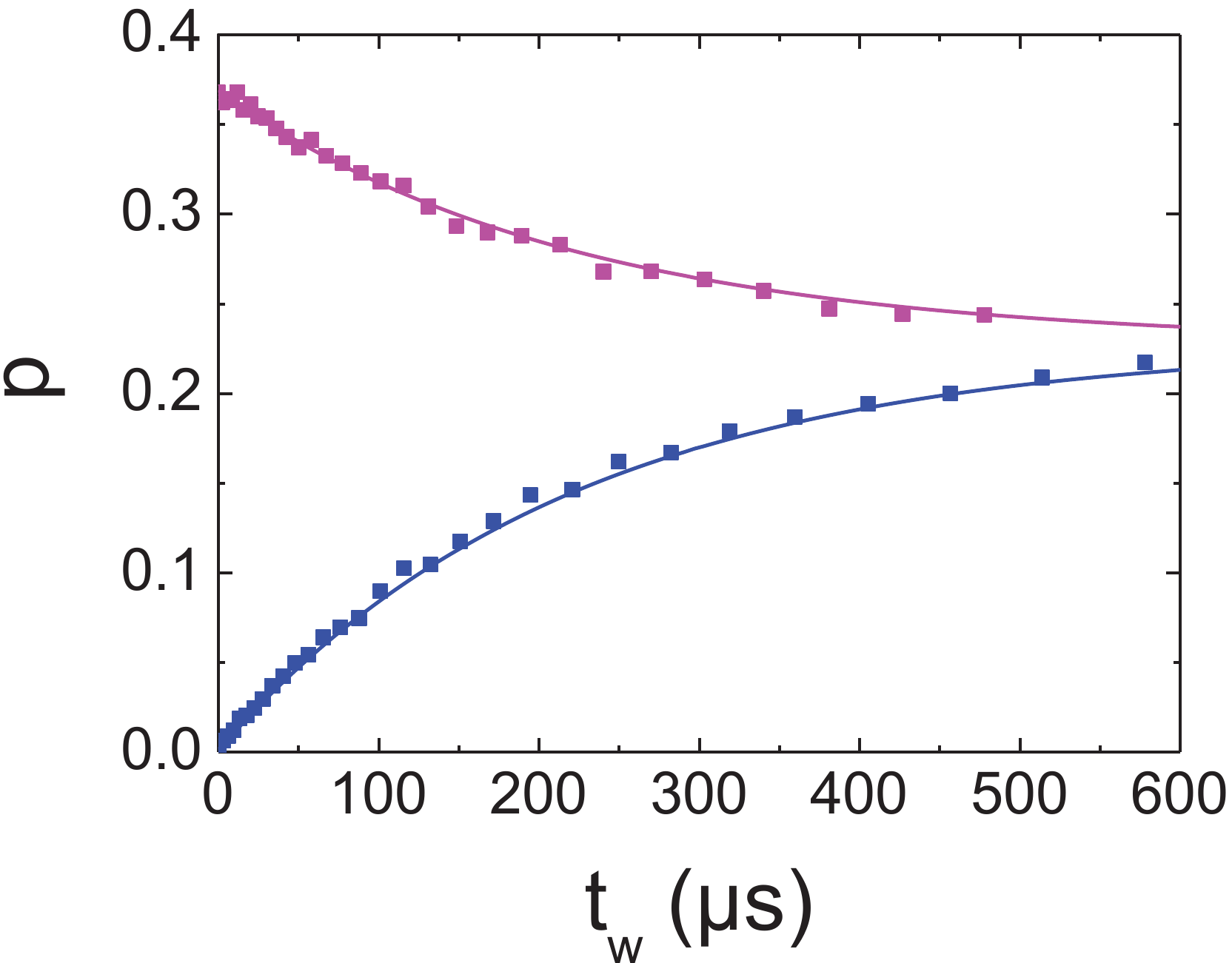} 
\par\end{centering}

\caption{Symbols: Poisoning probability as a function of time $t_{w}$ spent
at phase $\pi,$ for contact AC1 (\{1,0.07,0.07\}). Top curve is taken
with the pulse sequence shown in Fig.~\ref{Fluxpulse}, and shows
frequent poisoning ($p\left(0\right)=0.37$). Bottom curve has {}``antidote''
flux pulses after the current prepulse (see Fig.~\ref{clean pulse}),
which suppress initial poisoning ($p\left(0\right)=0$). Solid lines
are exponential curves with identical time constants $T_{1}=220\,\text{\textmu s}$
and asymptotic value $p=0.23$.}

\label{clean relax} 
\end{figure}

The experimental data indicate that when the phase $\delta$ across
the atomic contact is driven far from $\pi,$ the system quickly relaxes
to the ground state. Based on this result, we have developed a procedure
to get rid of trapped quasiparticles. The principle, illustrated in
Fig.~\ref{clean pulse}, is simply to sweep the flux phase over $2\pi,$
with a pair of symmetric {}``antidote'' pulses with amplitude $\pi$
each, hence insuring that whatever the starting phase, $\delta$ crosses
0 or $2\pi.$ Poisoning is cured by the antidote pulses, as shown
in Fig.~\ref{clean relax} where we compare relaxation curves with
and without them at $\varphi_{w}=\pi$ for contact AC1 (\{0.994,0.10,0.10\}),
the flux phase value where the poisoning probability was the highest.
Whereas data taken as in Fig.~\ref{relax} (right panel) show initial
poisoning with a probability $p_{0}=0.37,$ the poisoning probability
extrapolates to $0$ at $t_{w}=0$ when antidote pulses are applied
(Fig.~\ref{clean relax}, top curve), indicating that poisoning is
absent just after the pulses. Remarkably, the subsequent evolution
is identical in both situations, tending exponentially to $p=0.23$
with a time constant $T_{1}=220\,\text{\textmu s.}$ The identical
time constants and asymptotic poisoning probability was checked on
many atomic contacts. These data prove the robustness of the exponential
behavior, and that poisoning associated with residual quasiparticles
has the same effect whatever the initial configuration. %

\begin{figure}[h]
\begin{centering}
\includegraphics[clip,width=0.8\columnwidth]{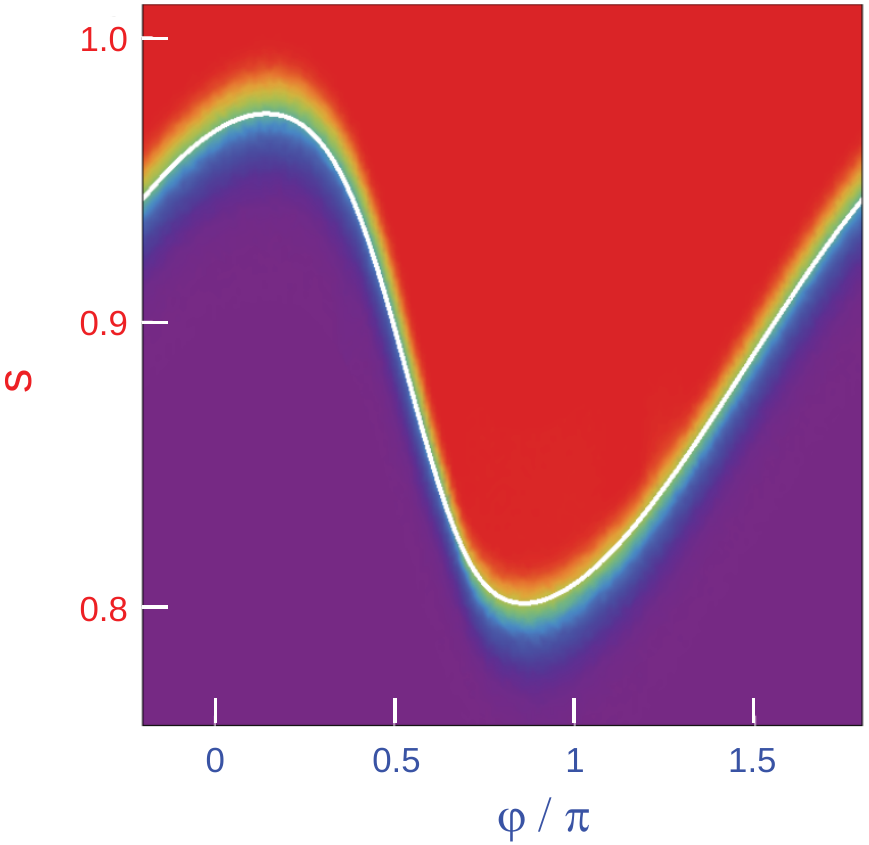} 
\par\end{centering}

\caption{Color plot of the switching probability $P_{sw}(s,\varphi)$ using
an antidote flux pulse 1~\textmu{}s before the measurement pulse,
and without prepulse, for the SQUID with contact AC0 with transmissions
\{0.95,0.45,0.10\}. The color scale is the same as for Fig.~\ref{Scurve}.
The white curve show the lines corresponding theoretically to $P_{sw}=0.5$
for the full contact. This plot is to be compared with Fig.~\ref{Scurve}.}

\label{clean3D} 
\end{figure}

Antidote pulses were employed to record $P_{sw}(s,\varphi)$ on the
SQUID with contact AC0, as shown in Fig.~\ref{clean3D}. We used
the simplest current bias pulses, shown in Fig.~\ref{pulses}, with
antidote pulses applied on the flux line 1~\textmu{}s before the
measurement pulse. The region displaying poisoning in Fig.~\ref{Scurve}
has completely disappeared, and switching is regular for all values
of the flux phase $\varphi.$%

The procedure presented here allows to obtain with very high probability
unpoisoned configurations, at least just after the antidote pulses.
Nevertheless, attempts to induce transitions from state $\left|0\right\rangle $
to state $\left|2\right\rangle $ with microwaves applied on the fast
flux line were unsuccessful in this experiment.

\section{Mechanism for initial poisoning\label{sec:Mechanism-for-intitial}}

The system always switches during the $0.1\,\text{\textmu s-}$long
prepulse, and a voltage of the order of $V=0.1\,\text{mV}$ develops
across the atomic contact. The phase $\delta$ then varies rapidly,
at the pace $V/\varphi_{0},$ and multiple Andreev reflections carry
a current across the atomic contact, which creates quasiparticles
on both sides of the contact\cite{chauvin}. For a contact of unit
transmission, two quasiparticles are being created for each turn of
the phase. The resulting number of quasiparticles is therefore $2\times0.1\,\text{\textmu s}\times0.1\,\mathrm{mV}\times2e/h\simeq10^{4}.$
When the bias current is reset to zero, the phase stops in a local
minimum of the global potential, and the configuration of the Andreev
levels freezes. The experiment indicates that after this step, the
population of the Andreev states generally does not correspond to
the equilibrium situation: there is a poisoning probability $p_{0}$
from which relaxation is subsequently observed. The exact process
of poisoning after switching is not understood presently. One can
however qualitatively grasp the mechanisms that can be relevant using
the picture of the {}``Andreev elevator'' that describes transport
in the finite voltage state, at least for small voltages\cite{chauvin}:
the phase varies linearly in time and the Andreev levels, in the semiconductor
representation, coincide periodically with the gap edges of the continuum
(when $\delta=2n\pi$), loading the lower level with a quasiparticle
at the lower continuum, and unloading it in the upper one. The quasiparticle
is sometimes transferred from the lower level to the upper one by
Landau-Zener tunneling. In this scenario, the system is always in
an even configuration. However, many quasiparticles are created in
the continuum, and the {}``loading'' (or {}``unloading'') process
can fail, leading to a situation where, after passing $\delta=2n\pi$,
the lowest state is unoccupied while the upper one has emptied (or
\emph{vice versa}), leading to an odd configuration. If the system
returns to the zero voltage state at this stage, the system stays
trapped in the odd configuration. The oscillating structure for $p_{0}\left(\varphi_{pp}\right)$
in Fig.~\ref{p0} suggests that interferences during the phase dynamics
may also play a role. Further work is clearly needed to clarify this
phenomenon.

\section{model for poisoning dynamics}

\begin{figure}[h]
\centering{}\includegraphics[width=1\columnwidth]{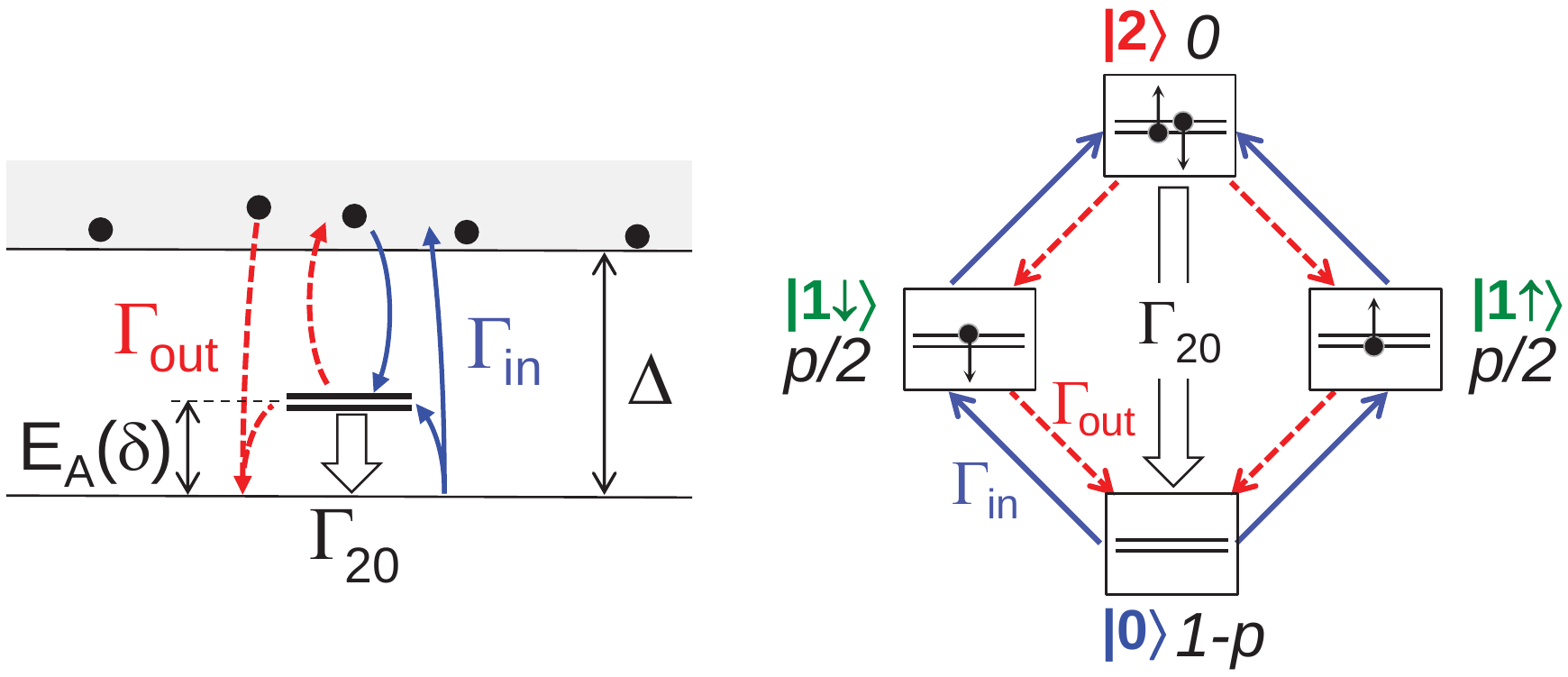}\caption{Model for the dynamics of the population of the Andreev configurations.
\textit{Left}: in the excitation representation. \textit{Right}: in
the configurations space, where the probability of the 4 configurations
are given in itallic. Quasiparticles can jump in and out the Andreev
levels at rates $\Gamma_{\mathrm{in}}$ and $\Gamma_{\mathrm{out}}.$
If both levels of the Andreev doublet are occupied, the system can
decay directly to the ground state (rate $\Gamma_{20}$).}
\label{model}
\end{figure}

We now discuss with the help of Fig.~\ref{model}, the details of
the model presented in Fig.$\,$4 of the Letter. Quasiparticles can
jump between the Andreev states and quasiparticle states in the continuum.%
The processes are in principle rather slow because they either require
energy absorption (rate $\Gamma_{out}$), or the presence of quasiparticles
in the continuum (rate $\Gamma_{in}$). Because the state with a double
excitation is never observed, we assume that the relaxation rate $\Gamma_{20}$
to the ground state is much larger than $\Gamma_{\mathrm{in}}$ and
$\Gamma_{\mathrm{out}}.$ This hypothesis is discussed in Appendix
B. 

The master equation for the populations of the ground state $\left|0\right\rangle $,
$1-p$, and of the odd configurations $\left|1\uparrow\right\rangle $
and $\left|1\downarrow\right\rangle $, $p/2$ each, is (see right
hand side of Fig.~\ref{model}):\begin{equation}
\frac{\mathrm{d}\left(1-p\right)}{\mathrm{d}t}=-2\left(1-p\right)\,\Gamma_{\mathrm{in}}+p\,(\Gamma_{\mathrm{out}}+\Gamma_{\mathrm{in}}),\end{equation}
the term $+p\Gamma_{\mathrm{in}}$ corresponding to transitions from
the odd configurations through the excited state $\left|2\right\rangle $
which relaxes very fast to the ground state. Assuming that $\Gamma_{\mathrm{in}}$
and $\Gamma_{\mathrm{out}}$ are time-independent, which is necessary
to obtain the observed exponential dependence of $p(t)$, one gets
equation (2) of the Letter. The assumption that $\Gamma_{\mathrm{in}},$
which is proportional to the quasiparticle density, is constant, means
that quasiparticles created in the voltage state after switching have
disappeared at the time scales probed by relaxation experiments. Experiments
with a succession of several prepulses indicate that this is the case.
One possibility is that they have already recombined. In the next
paragraph, we argue that they have probably already diffuse away.

To evaluate the decay of the density $n_{qp}$ of quasiparticles occurring
due to just diffusion, we start from the number of quasiparticles
created during the prepulse, $10^{4},$ as evaluated at the beginning
of Section \ref{sec:Mechanism-for-intitial}. Within the duration
of the prepulse, they spread over $\sim\sqrt{D_{qp}\times0.1\,\text{\textmu s}}>\sqrt{D_{N}\times0.1\,\text{\textmu s}}\simeq50\,\text{\textmu m}$
(with $D_{qp}=D_{N}/\left(1-\left(\Delta/E\right)^{2}\right)$ the
diffusion constant for quasiparticles of energy $E$ in the superconducting
state\cite{Diffusion constant}, $D_{N}$ the value in the normal
state), \emph{i.e.} over the whole area of SQUID (see Fig.~\ref{setup-2-1}(c)).
The density at the end of the prepulse is therefore $n_{qp}^{0}\approx100\,\mathrm{\mu m^{-3}}.$
Diffusion is then one-dimensional because it is limited by the 3 thin
connections (inductive lines), and a rough estimate (using again the
diffusion constant $D_{N}$ instead of $D_{qp}$) indicates that the
density decays within a few microseconds below $10\,\mathrm{\mu m^{-3}},$
which is the typical background value found for Al films at low temperatures
in Refs.~{[}\onlinecite{Shaw,Martinis}{]}. The quasiparticles
created by the prepulse can therefore be neglected when describing
the dynamics of poisoning relaxation.

\section{quasiparticle traps}

Let us note that in previous experiments with atomic contacts in SQUIDs\cite{chauvin,MLDR}
we didn't observe poisoning. In these works, the superconducting loop
was directly connected to very large normal metal electrodes, which
acted as efficient quasiparticle traps. In the present case, our SQUID
is contacted through very narrow and thin superconducting lines, and
clearly the small pieces of gold placed in direct contact with the
present SQUID body didn't act as efficient quasiparticle traps. In
fact, a normal metal electrode can only act as a trap if it is so
large that the diffusion time of a quasiparticle through it exceeds
the relaxation time of its energy.

On the other hand, the decoupling from dissipative parts of the circuit
is an important requirement\cite{Desposito} in order to explore the
physics of the Andreev qubit\cite{Lantz2002,Zazunov2003,Zazunov2005},
and the use of normal metal traps is probably not advisable. As a
consequence, there would always be some quasiparticles ready to get
trapped in the Andreev bound states. One method to get rid of them
is to use the {}``antidote pulses'' described in Section \ref{sec:Quasiparticle-poisoning-antidote}.
Another method is to have connecting wires with a lower gap than the
SQUID. We checked experimentally, using bilayers of Al and Cu, that
it allows to get rid of poisoning.

\section*{Appendix A: SQUID potentials}

\begin{figure}[h]
\begin{centering}
\includegraphics[clip,width=0.8\columnwidth]{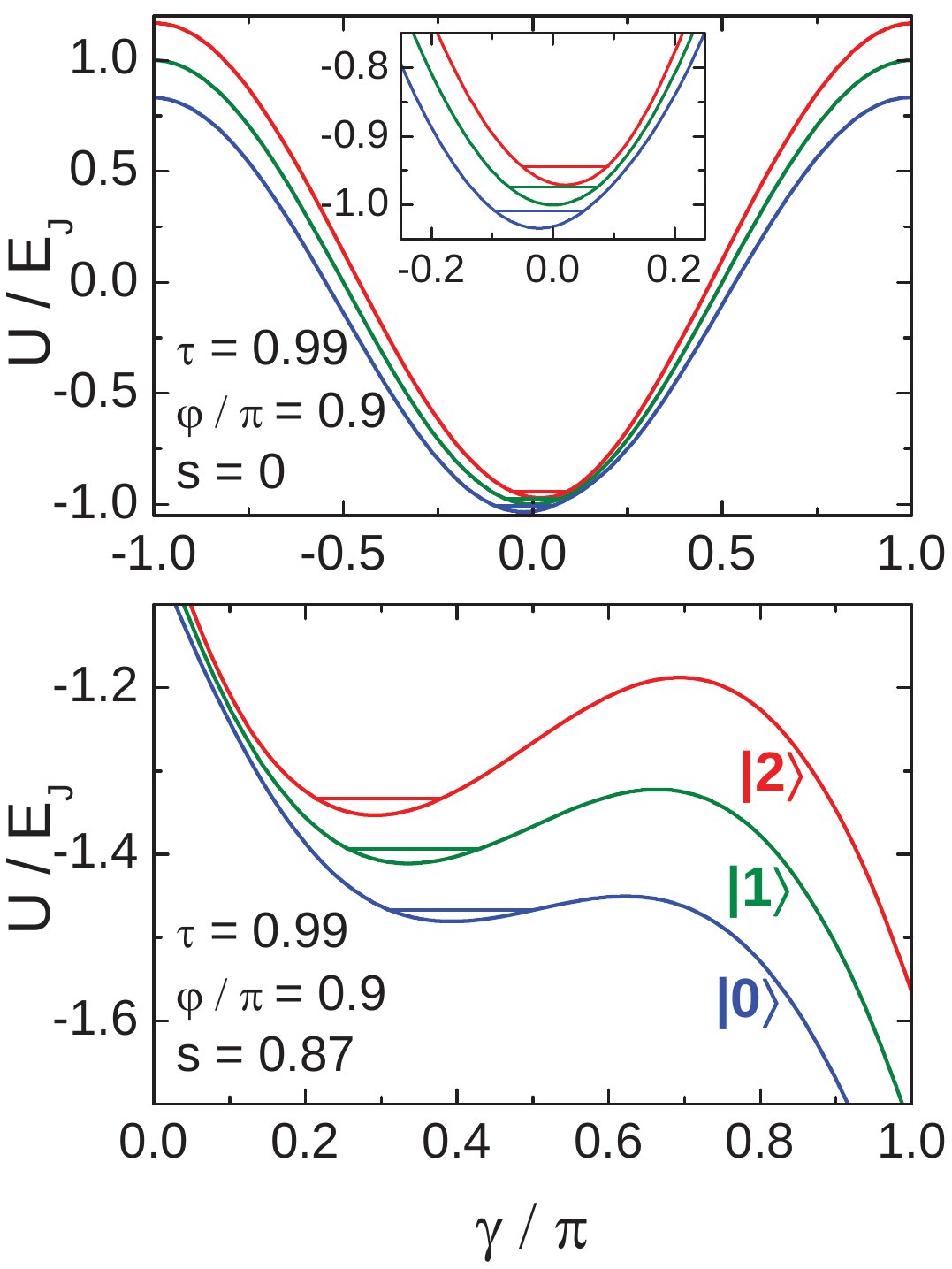} 
\par\end{centering}

\caption{Potentials of a SQUID with a single channel atomic contact with transmission
$\tau=0.99$, at $\varphi/\pi=0.9$ and $s=0$ (top panel) or $s=0.87$
(bottom panel). The three curves correspond, from top to bottom, to
the atomic contact in the ground state $\left|0\right\rangle $ (blue),
in the odd configuration $\left|1\right\rangle $ (green), and in
the excited state $\left|2\right\rangle $ (red). Horizontal lines
indicate the energy of the first level $\frac{1}{2}\hbar\omega_{p}$
in each potential, with the plasma frequency $\omega_{p}/2\pi$ calculated
semi-classically (at $s=0,$ $\omega_{p}^{\left|0\right\rangle }/2\pi=12.3\,\mathrm{GHz}$,
$\omega_{p}^{\left|1\right\rangle }/2\pi=14.1\,\mathrm{GHz}$, $\omega_{p}^{\left|2\right\rangle }/2\pi=14.8\,\mathrm{GHz}$;
at $s=0.87,$ $\omega_{p}^{\left|0\right\rangle }/2\pi=7.9\,\mathrm{GHz}$;
$\omega_{p}^{\left|1\right\rangle }/2\pi=9.7\,\mathrm{GHz}$; $\omega_{p}^{\left|2\right\rangle }/2\pi=10.9\,\mathrm{GHz}$).
The energy scale is $E_{J}=1.14\,\mathrm{meV\approx13\, k_{B}K.}$}

\label{potentials} 
\end{figure}

The SQUID potential is given by Eq.~(\ref{eq:potential}). If transitions
between configurations are slow compared to the plasma frequency,
which is the characteristic timescale for the dynamics in $U(\gamma),$
the escape rate is determined by the potential corresponding to the
instantaneous configuration\cite{average current}. As an example,
we plot in Fig.~\ref{potentials} the three potentials for a contact
with a single channel with transmission 0.99, at $\varphi/\pi=0.9$,
and for $s=0$ and $s=0.87.$ The plot at $s=0$ matters for the evaluation
of the transmission rates during $\Delta t,$ whereas the plot at
$s=0.87$ aims at illustrating why the switching rates are very different
in the ground state of the channel and in the odd configurations.
The energies $E_{J}=1.14\,\mathrm{meV}$ and $\Delta=0.193\,\text{meV}$
correspond to the parameters of the Josephson junction of the experiment.
The distance in energy between the potentials is the smallest when
$\delta=\pi,$ \emph{i.e. $\gamma=\pi-\varphi=0.1\pi$. }We also indicate
the ground state energy in each potential, as obtained from semi-classical
calculations. The quantum fluctutations of $\gamma$ (and hence of
$\delta$) are of the order of $0.1\pi.$ When the flux phase $\varphi$
is different from 0 and $\pi,$ as in the example of Fig.~\ref{potentials},
the three potentials do not have their minima at the same value of
$\gamma,$ an effect that is expected to modify the transition rates
between configurations\cite{These QLM}. 

At $s=0.87,$ the barrier heights are strongly reduced. The barrier
is the smallest in the ground state, and switching occurs only if
the system is in the ground state, and not in the odd configuration,
as is the case in Fig.~\ref{3Dplots} for $\varphi/\pi=0.9$ between
the dotted line and the solid line. %

\section*{Appendix B: relaxation between Andreev states}

\begin{figure}[h]
\begin{centering}
\includegraphics[clip,width=0.8\columnwidth]{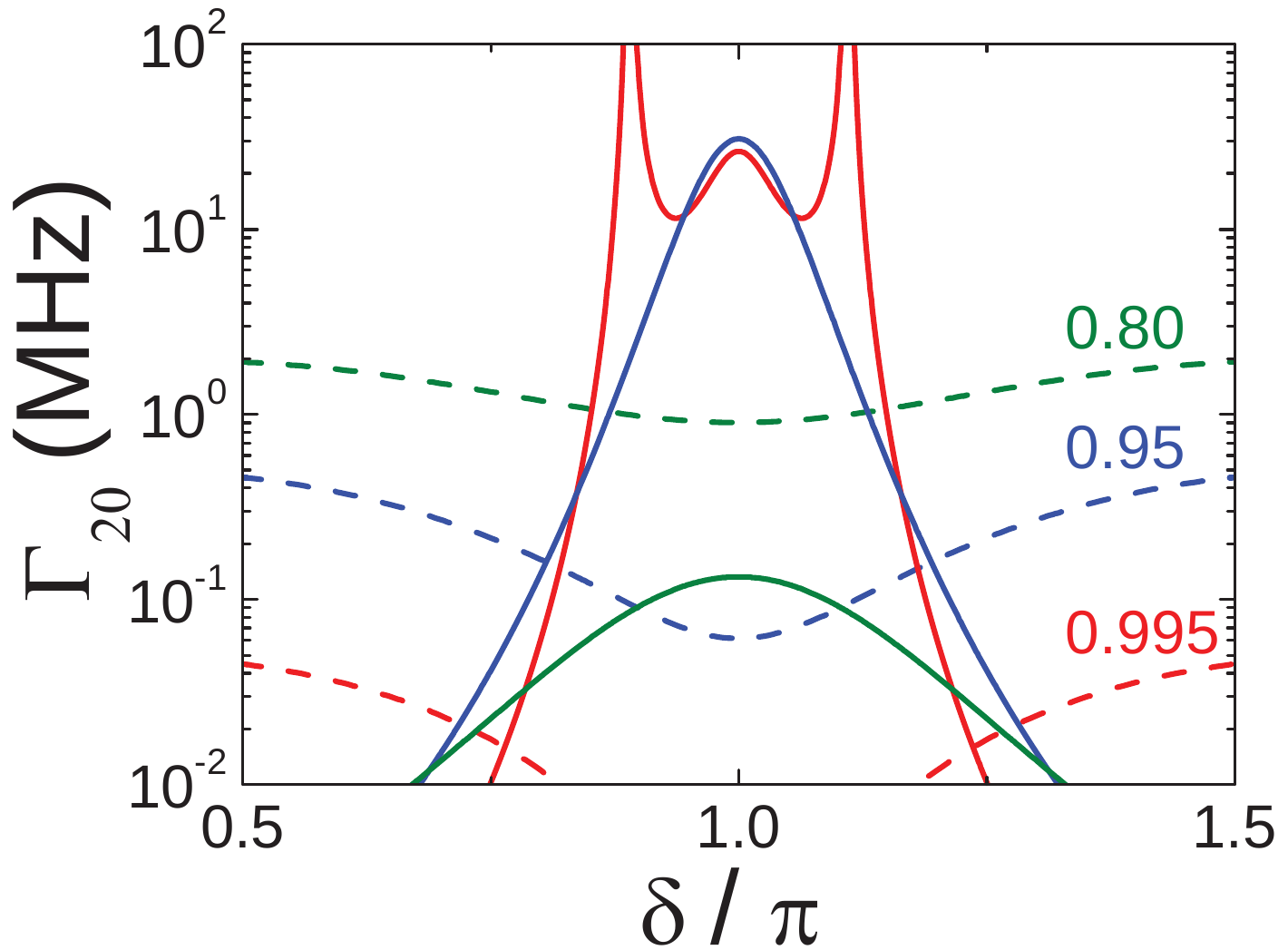} 
\par\end{centering}

\caption{Calculated prediction for the relaxation rate $\Gamma_{20}$ from
the excited singlet, for transmissions $\tau=0.8,$ 0.95 and 0.995.
Two contributions are plotted: rate for photon emission into the electronic
environment $\Gamma_{20}^{\mathrm{em}}$ (solid lines) and upper bound\cite{rateZazunov}
for the phonon emission rate $\Gamma_{20}^{\mathrm{ph}}$ (dashed
lines). For $\tau=0.995,$ three peaks are visible in $\Gamma_{20}^{\mathrm{em}},$
corresponding to the Andreev frequency $2E_{A}/h$ approaching the
low frequency mode of the electromagnetic environment at $0.4\,\mathrm{GHz}$
(central, broad peak) or coinciding with the high frequency mode at
$17\,\mathrm{GHz}$ (lateral, sharp peaks). For smaller transmissions,
only the effect of the high frequency mode is visible.}

\label{Gamma20} 
\end{figure}

The excited doublet $\left|2\right\rangle $ is never observed in
our experiments. We have attributed this fact to a large value of
the relaxation rate $\Gamma_{20}$ to the ground state. Two contributions
to this rate can be evaluated. The first one is the rate $\Gamma_{20}^{\mathrm{em}}$
of the process in which a photon is emitted into the electromagnetic
environment of the atomic contact\cite{Desposito}:\begin{equation}
\Gamma_{20}^{\mathrm{em}}=\frac{\pi\Delta}{\hbar}\frac{\mathrm{Re}\left[Z_{t}\left(2E_{A}/\hbar\right)\right]}{h/e^{2}}\frac{1-\tau}{\left(E_{A}/\Delta\right)^{3}}\left[\tau\,\sin^{2}\left(\frac{\delta}{2}\right)\right]^{2}\end{equation}
Note that this equation actually differs by the last factor from the
one in Refs.~{[}\onlinecite{Desposito}{]}, because we have used
to describe the Andreev system the hamiltonian proposed in Ref.~{[}\onlinecite{Zazunov2003}{]}
instead of the one introduced in Ref.~{[}\onlinecite{Ivanov}{]}
which does not enforces charge neutrality. For the actual circuit
design (Fig.~\ref{setup-2-1}), the environment impedance $Z_{t}$
is: \begin{equation}
Z_{t}^{-1}\left(\omega\right)=jC_{JJ}\omega+\frac{1}{jL_{JJ}\omega}+\frac{1}{jL\omega+\frac{1}{\left(r+\frac{1}{jC\omega}\right)^{-1}+R_{b}^{-1}}},\end{equation}
where $L_{JJ}=\varphi_{0}/I_{0},$ $C_{JJ}$ are the Josephson junction
inductance and capacitance, respectively. The eigenmodes of the two
LC circuits, at $1/(2\pi\sqrt{LC})\simeq0.5\,\mathrm{GHz}$ and $1/(2\pi\sqrt{L_{JJ}C_{JJ}})\simeq14.5\,\mathrm{GHz},$
couple to give two peaks near 0.4~GHz and 17~GHz; between 0.5~GHz
and 15.3~GHz and above 16.6~GHz, the impedance $\mathrm{Re}Z_{t}$
is smaller than $1\,\Omega$. The resulting rate $\Gamma_{20}^{\mathrm{em}},$
shown with solid lines in Fig.~\ref{Gamma20}, exceeds 1~MHz near
$\delta=\pi$ when the transmission is large enough. Note that these
estimations depends crucially on the value of the losses in the capacitor,
which are modelled by $r=0.5\,\Omega,$ a value determined from low-frequency
(0.5~GHz) measurements. When the Andreev frequency $2E_{A}/h$ is
large, the losses could easily be very different, and the rate much
larger.

The second contribution to $\Gamma_{20}$ is the rate $\Gamma_{20}^{\mathrm{ph}}$
of the process in which a phonon is emitted. An upper bound to this
rate\cite{rateZazunov} has been calculated in Refs.~{[}\onlinecite{Ivanov,Zazunov2005}{]}
in the limit $E_{A}\ll\Delta$:\begin{equation}
\Gamma_{20}^{\mathrm{ph}}\simeq\left(1-\tau\right)\frac{\Delta}{E_{A}}\tau_{ph}^{-1}\left(E_{A}\right),\end{equation}
with $\tau_{ph}^{-1}\left(E\right)=\kappa_{ph}\, E^{3}$ the bulk
electron-phonon relaxation rate at energy $E.$ Experiments\cite{Santhanam,Anthore,taueph}
on similar Al films give $\kappa_{ph}\simeq3\,\mathrm{\mu s^{-1}K^{-3}}$.
The corresponding predictions are shown with dashed lines in Fig.~\ref{Gamma20}
for transmissions 0.80, 0.95 and 0.995. Only for $\tau\leq0.80$ does
the rate $\Gamma_{20}^{\mathrm{ph}}$ exceed $1\,\mathrm{MHz}$ for
phases around $\pi.$ Therefore, the large value of $\Gamma_{20}$
in the phase interval where poisoning is observed (at the MHz scale,
which is the upper bound to the measured values of $\Gamma_{in,out}$,
see Fig.~4 of Letter) can be attributed to losses in the electromagnetic
environment of the atomic contact and, to a lesser amount, to phonon
emission. %

Given the values calculated for $\Gamma_{20}$, one could expect sharp
spectroscopic signatures of the transition $\left|0\right\rangle \rightarrow\left|2\right\rangle $
when microwaves are applied. However, no clear trace of this effect
could be found in the experiment. This could be due to imperfections
in the implementation of the microwave environment of the SQUID, or
to additional relaxation channels.